\documentclass[preprintnumbers,amsmath,amssymb,amsfonts,superscriptaddress,openany,nofootinbib]{revtex4}
\usepackage[utf8]{inputenc}
\usepackage{amsfonts}
\usepackage{amsbsy}
\usepackage{url}
\usepackage{enumerate}
\usepackage[justification=raggedright]{caption}
\usepackage{bbm}

\usepackage{graphicx,color}
\usepackage{amssymb,amsmath}
\usepackage{slashed}
\usepackage{hyperref}
\usepackage{braket}
\usepackage{simplewick}
\usepackage[title]{appendix}
\usepackage{caption}

\usepackage[subrefformat=parens]{subcaption}
\usepackage{ascmac}
\usepackage{ulem}
\usepackage{latexsym}
\usepackage{pifont}          
\usepackage{bm}
\usepackage{comment}

\newcommand{\n}{\nonumber}

\newcommand{\mr}[1]{\mathrm{#1}}

\newcommand{\h}[1]{\hspace{#1}}
\newcommand{\f}[2]{\frac{#1}{#2}}

\makeatletter
\@addtoreset{equation}{section}
\makeatother

\begin{document}
\preprint{YGHP-20-05, KUNS-2826}

\title{Topological structure of Nambu monopole in Two Higgs doublet models \\ -- Fiber bundle, Dirac's quantization and dyon --}

\author{Minoru~Eto}
\affiliation{Department of Physics, Yamagata University, Kojirakawa-machi 1-4-12, Yamagata, Yamagata 990-8560, Japan}
\affiliation{Research and Education Center for Natural Sciences, Keio University,
 4-1-1 Hiyoshi, Yokohama, Kanagawa 223-8521, Japan}

\author{Yu~Hamada}
\affiliation{Department of Physics, Kyoto University, Kitashirakawa, Kyoto 606-8502, Japan}

\author{Muneto~Nitta} 
\affiliation{Department of Physics, Keio University, 4-1-1 Hiyoshi, Kanagawa 223-8521, Japan}
\affiliation{Research and Education Center for Natural Sciences, Keio University, 4-1-1 Hiyoshi, Yokohama, Kanagawa 223-8521, Japan}

\begin{abstract}
We find a topologically non-trivial structure of the Nambu monopole in two Higgs doublet model (2HDM),
which is a magnetic monopole attached by two topologically stable $Z$ strings ($Z$ flux tubes) from two opposite sides.
The structure is in sharp contrast to the topological triviality of 
the Nambu monopole 
in the standard model (SM), which is attached by 
a single non-topological $Z$ string. 
It is found that
the Nambu monopole in 2HDM possesses 
the same fiber bundle structure with 
those of the `t Hooft-Polyakov monopole and the Wu-Yang description of the Dirac monopole,
as a result of the fact that 
the electromagnetic gauge field is well-defined even inside the strings and is non-trivially fibered around the monopole, 
while the Nambu monopole in the SM is topologically trivial 
because electroweak gauge symmetry is restored at the core of the string.
Consequently, the Nambu monopole in 2HDM can be regarded as an embedding of the 't Hooft-Polyakov monopole into the $SU(2)_W$ gauge symmetry, and the Dirac's quantization condition always holds,
which is absent for the Nambu monopole in the SM.
Furthermore, we construct a dyon configuration attached with the two strings.
\end{abstract}

\maketitle

\section{Introduction}
Magnetic monopoles have attracted great interests from many physicists since the seminal work by Dirac \cite{Dirac:1931kp},
which improves the asymmetry between electric and magnetic charges in the Maxwell equations 
and provide an explanation for the electric charge quantization.
While the monopole originally suggested by Dirac (Dirac monopole) is a singular point-like object 
accompanied with an infinitely-thin unphysical solenoid string (Dirac string),
it was reformulated by Wu and Yang \cite{Wu:1975es} from a viewpoint of fiber bundles without any string singularities.

In field theories,
magnetic monopoles often appear as topological solitons resulting from 
a non-trivial second homotopy group $\pi_2$ of the vacuum manifold,
which were first discovered by 't Hooft and Polyakov \cite{tHooft:1974kcl,Polyakov:1974ek} in the $SO(3)$ Georgi-Glashow model \cite{Georgi:1974sy}.
In such a model, the electric charge quantization can be understood by compactness of the unbroken $U(1)$ subgroup.
 In the same model, Julia and Zee \cite{Julia:1975ff} discovered a solution   
 with both the electric and magnetic charges, 
 called a dyon.
While the magnetic monopoles and dyons theoretically play crucial roles 
to understand non-perturbative aspects of (non-)supersymmetric field theories
\cite{Nambu:1974zg,Seiberg:1994aj,Seiberg:1994rs},
experimentally such monopoles have never been found in reality, 
except for condensed-matter analogues \cite{Castelnovo:2007qi,Ray:2014sga}.

A magnetic monopole configuration in the Standard Model (SM) was first considered by Nambu \cite{Nambu:1977ag},
which is called the Nambu monopole.
Since the Nambu monopole is attached by a vortex string,  
which is a physical string with finite thickness unlike the Dirac string for the Dirac monopole,
it is pulled by the tension of the string,
and cannot be stable.
The Nambu monopole nevertheless 
may be phenemonologically 
and cosmologically useful; for instance it is suggested to produce primordial magnetic fields before their disappearance
\cite{Vachaspati:2001nb,Poltis:2010yu}.
The electric charge quantization and the dyon associated with the Nambu monopole are also considered in Ref.~\cite{Vachaspati:1994xe}.
The mathematical reason of the instability is its trivial topology, that is,
the vacuum manifold of the SM is $S^3$ having a trivial second homotopy group $\pi_2$.
Likewise, the vacuum manifold $S^3$ has 
a trivial $\pi_0$ for domain walls and a trivial $\pi_1$ for cosmic strings.
Non-topological electroweak $Z$ strings (or magnetic $Z$-fluxes)
\cite{Vachaspati:1992fi,Vachaspati:1992jk,Achucarro:1999it,Brandenberger:1992ys,Barriola:1994ez,Eto:2012kb}
have been studied extensively, 
but they were shown to be unstable for the observed values of the Higgs mass $m_h\simeq 125~\mathrm{GeV}$ and the Weinberg angle $\sin^2\theta_W \simeq 0.23$
\cite{James:1992zp,James:1992wb}. 
The Nambu monopoles are the endpoints of these electroweak $Z$ strings \cite{Nambu:1977ag}.

On the other hand,
two Higgs doublet model (2HDM),
in which one more Higgs doublet is added to the SM, 
is one of the most popular extensions of the SM 
with a potential expectation to solve unsolved problems of the SM 
(for reviews, see, e.g., Refs.~\cite{Gunion:1989we,Branco:2011iw}).
Two Higgs doublet fields also appear in supersymmetric extension of the SM \cite{Nilles:1983ge,Haber:1984rc}.
In addition to the 125 GeV Higgs boson, 
it has four additional Higgs bosons,
which could be directly produced at the LHC, 
though there is no signal so far today, therefore placing lower bounds on masses of those additional scalar bosons.
Those lower bounds highly depend on parameter choices of the 2HDM. 
For more detailed phenomenological studies, see, e.g., Refs.~\cite{Trodden:1998ym,Kanemura:2015mxa,Kanemura:2014bqa,Kling:2016opi,Haller:2018nnx} and references therein.
One of the most remarkable aspects of 2HDM  
distinguishable from the SM may be
that it has a much richer vacuum structure than the SM,
thereby allowing a variety of 
topologically stable solitons, 
in addition to non-topological solitons \cite{La:1993je,Earnshaw:1993yu,Perivolaropoulos:1993gg,Bimonte:1994qh,Ivanov:2007de,Brihaye:2004tz,Grant:2001at,Grant:1998ci,Bachas:1996ap} analogous to the SM;
domain walls \cite{Battye:2011jj,Brawn:2011,Eto:2018tnk,Eto:2018hhg,Chen:2020soj,Battye:2020sxy}, 
membranes \cite{Bachas:1995ip,Riotto:1997dk}, 
and cosmic strings 
such as topological $Z$ strings \cite{Dvali:1993sg,Dvali:1994qf,Eto:2018tnk,Eto:2018hhg}
(see also Ref.~\cite{Bachas:1998bf}).
In particular, it was found 
\cite{Eto:2018tnk,Eto:2018hhg} that 
topological $Z$ strings in Refs.~\cite{Dvali:1993sg,Dvali:1994qf} 
are global strings confining non-Abelian fluxes in the cores, 
and are  
accompanied by non-Abelian moduli, 
analogous to non-Abelian strings in dense QCD
\cite{Balachandran:2005ev,Nakano:2007dr,Eto:2009kg,Eto:2009tr,Eto:2013hoa}.

In our previous papers \cite{Eto:2019hhf,Eto:2020hjb}, the present authors studied the Nambu monopole in 2HDM, 
which is a magnetic monopole attached with {\it two topological $Z$ strings from two opposite sides}.
The monopole is a regular solution of the EOM
and is topologically and dynamically stable 
when the Higgs potential has two global symmetries;
One is a global $U(1)$ symmetry that ensures the stability of the topological $Z$ strings.
The other is a discrete symmetry $\mathbb{Z}_2$ exchanging the topological $Z$ strings.
The string tensions pulling the monopole are balanced due to the $\mathbb{Z}_2$ symmetry,
and the monopole does not move unlike the Nambu monopole in the SM and can be regarded as a topologically stable $\mathbb{Z}_2$ kink on one string.
Once the  $\mathbb{Z}_2$ symmetry is explicitly broken, 
it starts to move along the string 
\cite{Eto:2020hjb}, 
and if the global $U(1)$ symmetry is broken, 
the string is attached by a domain wall 
\cite{Dvali:1993sg,Dvali:1994qf,Eto:2018tnk,Eto:2018hhg}.

One might wonder that the Nambu monopole in 2HDM is not a true magnetic monopole in several viewpoints;
First, its topological feature looks different from the ordinary magnetic monopole such as the 't Hooft-Polyakov monopole
because the monopole is not isolated but is attached with the two $Z$ string.
In other words, there is no non-trivial $\pi_2$ as in the SM.
Second, the magnetic charge of the monopole, $4 \pi \sin \theta_W/e$, is not an integer multiple of $2\pi/e$ with $e$ the electromagnetic coupling constant, 
and it seems inconsistent with the Dirac's quantization condition.

In this paper, we resolve the above two mysteries. 
We show that the Nambu monopole in 2HDM has the same topological structure with the 't Hooft-Polyakov monopole 
in the following sense,
while the Nambu monopole in the SM is topologically trivial.
We consider an infinitely large sphere $S^2$ surrounding the monopole,
on which the two $Z$ strings pass through,
and investigate the fiber bundle of the electromagnetic gauge field on the base space $S^2$.
It is found that the electromagnetic field is well-defined and regular everywhere on the sphere $S^2$ 
and is non-trivially fibered like the Wu-Yang description of the Dirac monopole, despite the trivial $\pi_2$.%
\footnote{
Note that a non-trivial $\pi_2$ is sufficient for a non-trivial fiber of the electromagnetic gauge field 
because of the mathematical relation $\pi_2(G/H)=\pi_1(H)$ where $G$ is a simple gauge group breaking into $H$.
Nevertheless, it is not necessary.
}
This is a remarkable difference from the Nambu monopole in the SM,
in which the electromagnetic gauge field cannot be defined at the center of the non-topological $Z$ string attached to the monopole  
since the electroweak gauge symmetry is restored there. 
It has a topologically trivial fiber bundle structure unlike the Wu-Yang's fiber bundle.
\footnote{
In Ref.~\cite{Vachaspati:1994xe}, it is insisted that the magnetic charge of the Nambu monopole in the SM is topologically protected, 
but we show that it is not the case.
Probably, the author of Ref.~\cite{Vachaspati:1994xe} did not pay enough attention to the fact that the electromagnetic gauge symmetry is not defined at the center of the $Z$ string due to the vanishing field value of the Higgs doublet.
}
Interestingly, the Nambu monopole in 2HDM can be regarded as an embedding of the 't Hooft-Polyakov monopole into the $SU(2)_W$ sector.
Furthermore, we derive the Dirac's quantization condition from the single-valuedness of wavefunctions around the non-trivially fibered electromagnetic field.
It is found that the condition always holds if the $Z$ flux is taken into acount as well, and thus there is no inconsistency.
We also consider a dyon configuration in 2HDM by describing a time-dependent ansatz 
and give a general quantization condition for the dyon charges as in Ref.~\cite{Vachaspati:1994xe}.

This paper is organized as follows.
In Sec.~\ref{154249_25Jun20} we will give a brief review on the 't Hooft-Polyakov monopole
and consider the fiber bundle and the Dirac's quantization condition.
In Sec.~\ref{061521_4Jun20} we will revisit the Nambu monopole in the SM contrasting its topological property to the 't Hooft-Polyakov case.
In Sec.~\ref{044315_30Jul20}, for later use, we will give a general definition for the electromagnetic gauge field and the field strengths in 2HDM.
Then we will consider the topological property of the electromagnetic gauge in Sec.~\ref{044348_30Jul20}.
The dyon configuration in 2HDM will be considered in Sec.~\ref{044406_30Jul20}.
Conclusion and discussion will be presented in Sec.~\ref{212143_21Jul20}.
In Appendix \ref{044424_30Jul20}, some useful relations are presented.
In Appendix \ref{030932_4Jun20}, background gauge condition is considered, which is used in Sec.~\ref{044406_30Jul20}.
In Appendix \ref{044644_30Jul20}, a description in the singular gauge for the Nambu monopole in 2HDM is presented.

\section{Review on 't Hooft-Polyakov monopole}
\label{154249_25Jun20}
\subsection{The 't Hooft-Polyakov monopole}
In this section, 
we breifly review the 't Hooft-Polyakov monopole 
with a particular emphasis on its topological property.
For more details, see, e.g., Ref.~\cite{Balian:2005joa}.
We start with the Yang-Mills-Higgs model with $SU(2)$ gauge group:
\begin{equation}
 \mathcal{L}= \f{-1}{4} W_{\mu\nu}^a W^{\mu\nu a} + (D_\mu \Phi)^a (D^\mu \Phi)^a -V(\Phi)
\end{equation}
with a potential term
\begin{equation}
 V(\Phi)= -m_\Phi^2 (\Phi^a)^2 + \lambda ((\Phi^a)^2)^2,
\end{equation}
where $\Phi^a$ ($a=1,2,3$) is the Higgs field in the adjoint representation%
\footnote{Note that the true gauge group is $SU(2)/\mathbb{Z}_2$ under the presence of the adjoint representation only.
Nevertheless, we will use $SU(2)$ which is the universal covering of $SO(3)$ for ease of simple notations.}
,
$W_\mu^a$ is the gauge field 
with the field strength
$W_{\mu\nu}^a \equiv \partial_\mu W_\nu^a - \partial_\nu W_\mu^a - g \epsilon^{abc} W_\mu^b W_\nu^c $.
The Higgs vacuum is given by
\begin{equation}
 \Phi_0^a = (0,0,v)^t,\label{151027_10Jun20}
\end{equation}
with $v \equiv m_\Phi /\sqrt{2 \lambda}$.
The vacuum \eqref{151027_10Jun20} spontaneously breaks $SU(2)$ into $U(1)$.

Let us consider the 't Hooft-Polyakov monopole located at the origin.
In this paper, it is enough to consider an asymptotic configuration describing the monopole at large distances $r\to \infty$, given by 
\begin{equation}
\Phi^a(\theta,\varphi)\bigg|_{\infty} = v \f{x^a} {r}, \quad g\, W_{i}^a(\theta,\varphi)\bigg|_{\infty} = \epsilon^{iab}\f{x^b}{r^2}\label{031411_1Apr20} ,
\end{equation}
where $\theta$ and $\varphi$ are the zenith and azimuth angles, respectively.
This is sometimes called the hedgehog gauge.
In this ansatz, an unbroken gauge group is determined by a unit vector $n^a$ defined by
\begin{equation}
 n^a \equiv \f{\Phi ^a}{|\Phi|}\label{181300_26Jun20}
\end{equation}
with $|\Phi| \equiv \sqrt{\Phi^a \Phi^a}$.
The vector $n^a$ is regular everywhere except for a region in which $|\Phi|=0$ (corresponding to the center of the monopole).
We denote the unbroken gauge group by $U(1)_\mr{EM}$.
Indeed, the Higgs field in Eq.~\eqref{031411_1Apr20} is invariant under the $U(1)_\mr{EM}$ rotation:
\begin{equation}
 e^{-i \alpha  n^a \sigma^a} \Phi e^{i \alpha n^b \sigma^b } = \Phi
\end{equation}
with $\Phi = \Phi^a \sigma^a/2$, and $\alpha$ an arbitrary constant.
From Eq.~\eqref{031411_1Apr20}, $n^a$ has a hedgehog structure at large distances, $n^a \sim x^a/r$,
and hence is a map from the two dimensional sphere $S^2$ spanned by $(\theta,\varphi)$ at spatial infinity 
into a two dimensional internal sphere $S^2 \simeq \{n^a |~ (n^a)^2=1\} $ with a unit winding number:
\begin{equation}
\f{1}{8\pi} \int_{S^2} dS_{k}\epsilon^{ijk}\epsilon^{abc} n^a \partial_i n^b \partial_j n^c =1,\label{104517_30Jul20}
\end{equation}
where the integration is performed on the two-dimensional sphere surrounding the monopole at the spatial infinity.
In addition, it satisfies at large distances the following identity:
\begin{equation}
 \left(D_\mu n\right)^a = \partial_\mu n^a - g \epsilon^{abc} W_\mu^b n^c =0.\label{215910_11Jun20}
\end{equation}

The electromagnetic field strength is defined by a projection of the $SU(2)$ field strength onto the $U(1)_\mr{EM}$ subgroup,
\begin{align}
 F_{ij}^\mr{EM} &\equiv W_{ij}^a n^a\label{053024_1Apr20} \\
 &\sim \f{1}{g}\epsilon_{ijk} \f{x^k}{r^3} \quad (\mathrm{for~}r\to \infty)\label{045043_1Apr20}
\end{align}
and the magnetic field is
\begin{equation}
 B_i \equiv \f{1}{2} \epsilon_{ijk}F^\mr{EM}_{jk} \sim \f{1}{g} \f{x^i}{r^3}.
\end{equation}
Therefore, the ansatz \eqref{031411_1Apr20} describes a magnetic monopole with a magnetic charge $q_M = \int_{S^2}dS_i B_i = 4 \pi /g$.
Note that, if one defines the electromagnetic gauge field $A_\mu$ as
\begin{equation}
 A_\mu \equiv W_\mu^a n^a,
\end{equation}
then Eq.~\eqref{053024_1Apr20} can be rewritten as
\begin{equation}
 F_{ij}^\mr{EM} = \partial_i A_j - \partial_j A_i + \f{1}{g } \epsilon ^{abc} n^a \partial_i n^b \partial_j n^c,\label{050120_1Apr20}
\end{equation}
where we have used Eq.~\eqref{215910_11Jun20}.
Thus, the electromagnetic field strength is composed of
a contribution purely from the electromagnetic gauge field $A_\mu$ and 
an additional contribution from the Higgs field $\Phi$,
which explains why there can be a non-zero divergence of the magnetic flux.

We can consider a more general case that the magnetic charge is larger than $4\pi / g$.
This is achieved by considering an ansatz for $\Phi^a$ or $n^a$ with a general winding number, {\it i.e.},
\begin{equation}
\f{1}{8\pi} \int_{S^2} dS_{k}\epsilon^{ijk}\epsilon^{abc} n^a \partial_i n^b \partial_j n^c = m
\label{232201_12Jun20}
\end{equation}
where $m$ must be an integer since it is a topological number.
In other words, such a configuration is in a topological sector labeled by $m\in \mathbb{Z}$ of the second homotopy group 
$\pi_2[SU(2)/U(1)] \simeq \pi_2(S^2)= \mathbb{Z}$.
The magnetic charge is obtained by integrating Eq.~\eqref{050120_1Apr20} on $S^2$,
\begin{equation}
q_M= \int_{S^2} dS_k \epsilon ^{ijk} F_{ij}^\mr{EM} = \f{4\pi}{g} m,
\end{equation}
where we have used a fact that the integration for $A_i$ vanishes as long as $A_i$ is regular.
Thus $q_M$ is quantized with the topological number $m$.

\subsection{Singular gauge for the 't Hooft-Polyakov monopole}
It is instructive to see the 't Hooft-Polyakov monopole in a singular gauge (also called the string gauge).
Firstly, let us consider the single monopole configuration with $m=1$ in Eq.~\eqref{031411_1Apr20}.
To move from the hedgehog gauge to the string gauge, we introduce a gauge transformation $U$ such that
\begin{equation}
 n^a \f{\sigma ^a}{2} \to  (n')^a \f{\sigma ^a}{2} = U^\dagger n^a \f{\sigma ^a}{2} U = \f{\sigma ^3}{2},
\end{equation}
or equivalently $ n^a = (\sin\theta\cos\varphi,\sin\theta\sin\varphi,\cos\theta)^t \to (n')^a = (0,0,1)^t$.
Such a transformation $U$ is given by
\begin{align}
 U (\theta , \varphi) &= e^{-i \sigma ^3 \varphi / 2} e^{-i \sigma ^2 \theta / 2} e^{i \sigma ^3 \varphi / 2}
 = \begin{pmatrix}
     \cos \f{\theta}{2} & -\sin \f{\theta}{2} e^{-i \varphi} \\
     \sin \f{\theta}{2} e^{i \varphi}  & \cos \f{\theta}{2}
    \end{pmatrix},\label{044222_1Apr20}
\end{align}
where we should note that $U$ is a singular function on $\theta=\pi$ (the negative side of the $z$-axis) but not on $\theta=0$.

By acting the singular transformation $U$ on Eq.~\eqref{031411_1Apr20}, we obtain, at large distances,
\begin{equation}
 \Phi ^a \to  \Phi ^a_0 =(0,0,v)^t,
\end{equation}
\begin{align}
 W_i &\to W_i^\mr{sing.}= U ^\dagger \left(W_i -\f{i}{g} \partial_i\right) U 
 = \f{1}{2gr} \f{1- \cos \theta}{\sin \theta} \hat{\varphi}_i \sigma^3,\label{202504_10Jun20}
\end{align}
with $W_i \equiv W_i^a \f{\sigma^a}{2}$ and  $\hat{\varphi}_i = (-\sin \varphi, \cos \varphi,0)^t$.
The transformed gauge field $W_i^\mr{sing.}$ has a line singularity at $\theta = \pi$, 
which comes from the derivative
$U^\dagger \partial_i U$,
and hence this gauge is dubbed the singular gauge.
In this gauge, the Higgs field is constant (at large distances),
and thus the electromagnetic gauge field is defined straightforwardly 
as the $\sigma^3$-direction of the $SU(2)$ gauge field,
\begin{align}
 A_i = \mr{tr}\left[\sigma^3 \, W_i^\mr{sing.}\right] 
 &= \f{1}{gr} \f{1- \cos \theta}{\sin \theta} \hat{\varphi}_i 
 =  \f{1}{g} (1- \cos \theta) \partial_i\varphi. 
 \label{051637_1Apr20}
\end{align}

Let us calculate the electromagnetic field strength Eq.~\eqref{050120_1Apr20}, 
which is a gauge invariant quantity and should match with the result \eqref{045043_1Apr20}.
The third term in Eq.~\eqref{050120_1Apr20} would seem to vanish naively in this gauge.
However, this is not true 
because we have to take into account the singularity from the derivative $\partial_i \varphi$.
Indeed, under the transformation $U$, the third term transforms as
\begin{equation}
 \f{1}{g } \epsilon ^{abc} n^a \partial_i n^b \partial_j n^c \to -\f{4 \pi}{g}\epsilon_{ij} \theta(-z) \delta(x) \delta(y),
\end{equation}
which describes an infinitely thin solenoid on $\theta=\pi$ (terminating at the origin), similarly to a Dirac string for a Dirac monopole.
This line singularity is completely canceled by that from the gauge field \eqref{051637_1Apr20},
and thus we obtain the field strength as
\begin{align}
 F_{ij}^\mr{EM} &= \left(\f{1}{g} \sin \theta \partial_{[i} \theta \,\partial_{j]}\varphi + \f{1}{g} (1- \cos \theta) \partial_{[i}\partial_{j]}\varphi \right)
 -\f{4 \pi}{g}\epsilon_{ij} \theta(-z) \delta(x) \delta(y) \nonumber \\
&= \f{1}{g}\epsilon_{ijk}\f{x^k}{r^3},
\end{align}
which is regular, coinciding with Eq.~\eqref{045043_1Apr20}.
Therefore, this configuration is an embedding of the Dirac monopole (divergence-less Dirac potential + Dirac string) into the $SU(2)$ gauge fields. 

The singularity in $A_i$ is gauge artifact in the sense that it does not originally exist in the hedgehog gauge.
Thus, it should not be observed by a test particle with an electric charge $e=g/2$
(spin-$1/2$ representation of $SU(2)$).
We thus obtain a condition that the Aharonov-Bohm (AB) effect from the singularity is not observed by the test particle:
\begin{equation}
 e \times q_M = \f{g}{2} q_M = 2\pi n \h{2em} n \in \mathbb{Z} \label{055759_1Apr20} ,
\end{equation}
which is nothing but the Dirac's quantization condition.
Recalling $q_M=4 \pi /g$, this condition is automatically satisfied with $n=1$.

We consider a more general case with a multiple topological number $m>1$ in Eq.~\eqref{232201_12Jun20},
which provides a magnetic charge $q_M = 4\pi m / g$.
In the singular gauge, the electromagnetic gauge field producing the magnetic charge is given by
\begin{equation}
 A_i =  \f{m}{g} (1- \cos \theta) \partial_i\varphi ,
\end{equation}
which has also a singularity on $\theta= \pi$.
The condition \eqref{055759_1Apr20} leads to
\begin{equation}
  2\pi m =  2\pi n \h{2em} n \in \mathbb{Z} ,
\end{equation}
which always holds with $m=n$.
Therefore, the Dirac string appears in the singular gauge of the 't Hooft-Polyakov monopole,
but it is always unobserved because the Dirac quantization condition is ensured by the topological quantization for $\pi_2(S^2)$.

\subsection{Wu-Yang monopole bundle}
A fiber bundle is a topological space consisting of a base space $B$ and a fiber $F$.
The fiber bundle is locally homeomorphic to the direct product of the base space and the fiber: $B \times F$,
but is not globally in general.
A famous example is the M\"obius strip,
which is a non-trivial fiber bundle consisting of a circle as a base space and a segment as a fiber.
It seems locally as a cylinder, but it has an overall twist, 
which is visible only globally.
In order to study global structures of fiber bundles, 
it is often useful to divide the base space into several patches
such that the fiber bundle is homeomorphic to the direct product on each patch.
A non-trivial structure appears after gluing the patches in a non-trivial way.
For the M\"obius strip, we introduce two segments $[0,2\pi]\times [-1,1]$ and glue the two ends $0$ and $2\pi$
in such a way that $(0,x)\in [0,2\pi]\times [-1,1] $ is identified with $(2\pi,-x)\in [0,2\pi]\times [-1,1]$.
It has a non-trivial global twist (or $\mathbb{Z}_2$ action: $x \to -x$) for the segment $[-1,1]$.

We here discuss the Dirac quantization condition from the viewpoint of the fiber bundle.
That provides us a clear connection with the Wu-Yang description of the Dirac monopole \cite{Wu:1975es},
in which it is decribed in two patches and there is no line singularity (Dirac string) we have seen above.
Let us consider again the single monopole configuration in  Eq.~\eqref{031411_1Apr20}.
We introduce a two-dimensional sphere $S^2$ surrounding the monopole at spatial infinity
parametrized by the azimuthal angle $\varphi\in [0, 2\pi)$ and the zenith angle $\theta \in [0,\pi]$.
Recall that $|\Phi|^2 = v^2$ and $n^a= x^a /r $ on the sphere.
To define the electromagnetic gauge field globally on the sphere {\it without} any singularities,
we divide it into two patches: hemi-spheres $R_N$ and $R_S$, 
\begin{equation}
  R^N : 0 \leq \theta \leq \f{\pi}{2} , \h{2em}   R^S : \f{\pi}{2} \leq \theta \leq \pi ,
\end{equation}
which have an overlap region on the equator $\theta = \pi /2$,
and we introduce $SU(2)$ gauge transformations defined on each hemi-sphere:
\begin{align}
R^N : ~ U ^N(\theta , \varphi) &= e^{-i \sigma ^3 \varphi / 2} e^{-i \sigma ^2 \theta / 2} e^{i \sigma ^3 \varphi / 2}
= \begin{pmatrix}
     \cos \f{\theta}{2} & -\sin \f{\theta}{2} e^{-i \varphi} \\
     \sin \f{\theta}{2} e^{i \varphi}  & \cos \f{\theta}{2}
 \end{pmatrix} \\
R^S : ~ U^S (\theta , \varphi) &= e^{-i \sigma ^3 \varphi / 2} e^{-i \sigma ^2 (\pi-\theta) / 2} e^{i \sigma ^3 \varphi / 2} (i \sigma^2)
= \begin{pmatrix}
    e^{-i \varphi} \cos \f{\theta}{2} & -\sin \f{\theta}{2}  \\
     \sin \f{\theta}{2}  &  e^{i \varphi} \cos \f{\theta}{2}
 \end{pmatrix} .
\end{align}
The both transformations bring the hedgehog-like vector $n^a= x^a/r$ into the uniform vector on each region as
\begin{align}
 U^N : ~ n^a\f{\sigma^a}{2} &\to \left(U^N\right)^\dagger n^a\f{\sigma^a}{2} U^N= \f{\sigma^3}{2}\label{215751_30Apr20}\\
 U^S : ~ n^a\f{\sigma^a}{2} &\to \left(U^S\right)^\dagger n^a\f{\sigma^a}{2} U^S= \f{\sigma^3}{2},\label{215757_30Apr20}
\end{align}
or equivalently, $\Phi^a \to \Phi_0^a=(0,0,v)$ in both $R_N$ and $R_S$.
Note that $U^N$ and $U^S$ are regular on each hemi-sphere $R^N$ and $R^S$.
On the two hemispheres, the unbroken gauge group $U(1)_{\rm EM}$ is \textit{uniformly} defined as the $\sigma^3$ subgroup of $SU(2)$.

Since we have $U^N \Phi = U^S \Phi = (0,0,v)^t$
on the overlap region of the two patches, $\theta = \pi/2$,
the transition function $(U^S)^{-1} U^N $ must leave $\Phi$ invariant,
{\it i.e.}, $(U^S)^{-1} U^N \in U(1)_{\rm EM}$.
Indeed, we have
\begin{align}
 \left(U^S\right)^{-1} U^N= e^{ i \varphi \sigma^3},\label{000509_1May20}
\end{align}
and thus it does not change $\Phi^a_0 = (0,0,v)^t$.
The transition function is a map from the equator $S^1:\varphi \in [0,2\pi)$ into $U(1)_{\rm EM}(\simeq S^1)$ with a winding number unity,
and hence the two patches are glued in a topologically non-trivial way.

Let us define the electromagnetic gauge field after the transformations in Eqs.~\eqref{215751_30Apr20} and \eqref{215757_30Apr20} on each patch $R^N$ and $R^S$ as
\begin{align}
 A_i^N \equiv W_i^3 &= \f{1}{g} \f{1- \cos \theta}{r\sin \theta} \hat{\varphi}_i\label{032954_11Jun20} \\
 A_i^S \equiv W_i^3 &= -\f{1}{g} \f{1+ \cos \theta}{r\sin \theta} \hat{\varphi}_i.\label{032959_11Jun20}
\end{align}
Importantly, they are regular on each patch as desired.
On the equator $\theta= \pi/2$, they differ by the $U(1)_{\rm EM}$ gauge transformation:
\begin{equation}
 A_i ^N 
=A_i ^S + \f{2}{gr} \hat{\varphi}_i \equiv A_i ^S + \Delta A_i.
\end{equation}
The difference $\Delta A_i$ originates from the transition function \eqref{000509_1May20} as
\begin{equation}
\Delta A_i \f{\sigma^3}{2}= \f{-i}{g} \left[\left( U^S\right)^{-1} U^N \right] ^\dagger \partial_i \left[\left( U^S\right)^{-1} U^N\right].
\end{equation}

We introduce the first Chern number, 
which characterizes the topology of the fiber bundle with the fiber $U(1)_\mr{EM}$ over the base space $S^2$ surrounding the monopole,
as
\begin{equation}
  \int_{S^2} c_1 \equiv \f{e}{2 \pi} \int _{S^2} F^\mr{EM} ,\label{033012_11Jun20}
\end{equation}
where $c_1$ is called the first Chern class and $S^2$ consists of $R^N$ and $R^S$.
The magnetic charge of the monopole coincides with the first Chern number up to an overall constant.
Substituting Eqs.~\eqref{032954_11Jun20} and \eqref{032959_11Jun20} into Eq.~\eqref{033012_11Jun20},
we have
\begin{align}
 \int_{S^2} c_1 & = \f{g}{4 \pi} \int _{S^1: \, \theta =\pi/2} \left(A^N - A^S\right)  = \f{g}{4 \pi} \int _{S^1} \Delta A\label{033311_11Jun20} \\
& = \f{1}{2 \pi} \int_0 ^{2\pi} d \varphi =1,\label{011144_30Apr20}
\end{align}
where $F^\mr{EM}\equiv \frac{1}{2} F_{ij}^\mr{EM} dx^i \wedge dx^j$, $A^{N(S)} \equiv A_i ^{N(S)} dx^i $ 
and we have used the definition of the electric charge $e \equiv g/2$ and Stokes theorem.
From Eq.~\eqref{033311_11Jun20}, the first Chern number counts the winding number of the transition function on the equator,
and the nonzero value means that the two patches $R^N$ and $R^S$ are non-trivially glued.
Thus, the transition function is in a non-trivial topological sector classified by the first homotopy group $\pi_1(S^1)=\mathbb{Z}$.
In the language of fiber bundles, 
this means that the topological space consisting of the base space $S^2$ and the fiber $U(1)_{\rm EM}$ is non-trivial 
(not $S^2 \times U(1)_{\rm EM}$) 
and is homeomorphic to $S^3$ (the Hopf fibration).
Therefore, the configuration in Eqs.~\eqref{032954_11Jun20} and \eqref{032959_11Jun20} is regarded as an embedding of the Wu-Yang description of the Dirac monopole into the $SU(2)$ gauge theory.

In this formulation, the Dirac's quantization condition can be expressed as single-valuedness of a wave function of a test particle with an electric charge $e$.
When the test particle goes once around the equator $\theta=\pi/2$,
it receives the AB phase
\begin{equation}
 \theta_{AB} = e \oint dx_i A_i.
\end{equation}
In order for the wave function to be single-valued, the two effects calculated on the two patches $R^N$ and $R^S$ must be equivalent.
Thus, we obtain the following condition:
\begin{align}
 & e \oint_{\theta = \pi/2} dx_i \Delta A_i = 2 \pi n ~ (n \in \mathbb{Z}) \\
 \Leftrightarrow &~ 2 \pi  \int_{S^2} c_1 = 2 \pi n,
\end{align}
which is automatically satisfied with $n=1$ by Eq.~\eqref{011144_30Apr20}.
This result can be extended to more general cases with monopole charges larger than unity.

In this section, we have seen the 't Hooft-Polyakov monopole in two ways, in the singular gauge and in the language of fiber bundles.
While the former corresponds to the Dirac monopole and the Dirac string, the latter is related with the Wu-Yang description without the Dirac string.
Both of them provide the Dirac's quantization condition but it is automatically satisfied due to the non-trivial second or first homotopy groups, $\pi_2(G/H)= \mathbb{Z}$ or $\pi_1(S^1)= \mathbb{Z}$.
In the following sections, we study these for a Nambu monoole
in the SM and 2HDM.

\section{Nambu monopole in the Standard Model}
\label{061521_4Jun20}
\subsection{Hedgehog gauge}
We here discuss the Nambu monopole \cite{Nambu:1977ag} in the SM  in the hedgehog gauge.
The gauge-Higgs sector of the SM Lagrangian is 
\begin{align}
 \mathcal{L} &= -\f{1}{4}W_{\mu\nu}^a W ^{a\mu\nu} - \f{1}{4} Y_{\mu\nu} Y^{\mu\nu} + | D_\mu \Phi_{\mathrm{SM}} |^2 - \lambda\left( |\Phi_{\mathrm{SM}} |^2 - v^2\right)^2 ,
\end{align}
where $\Phi_{\mathrm{SM}}$ is the SM Higgs doublet.
The covariant derivative is $D_\mu \Phi_\mr{SM}=(\partial_\mu - i\f{g}{2} W_\mu^a \sigma^a - i\f{g'}{2}Y_\mu)\Phi_\mr{SM}$ where
$W$ and $Y$ are the $SU(2)_W$ and $U(1)_Y$ gauge fields, respectively.

The Nambu monopole, which is a magnetic monopole attached by a $Z$ string ($Z$-flux tube), 
was first given by Nambu \cite{Nambu:1977ag}.
Since the electroweak symmetry breaking $SU(2)_W \times U(1)_Y \to U(1)_{\rm EM}$ is topologically trivial,
neither the Nambu monopole nor the $Z$ string can be topologically stable.
We consider a situation that 
the monopole lies on the origin and the $Z$ string 
emanating from it
is put on the positive side of the $z$ axis ($\theta = 0$).
At large distances from the monopole, $r\to \infty$, 
the configuration describing the monopole is given by
\begin{eqnarray}
 \Phi_{\mathrm{SM}} &=& v \begin{pmatrix}
  e^{ i\varphi} f(\theta)\cos \f{\theta}{2} \\  \sin \f{\theta}{2}
    \end{pmatrix},\\
 g W_i ^{a
 } &=& -\cos^2 \theta_W \f{x^a}{r} h(\theta)(1+\cos \theta) \partial_i \varphi -\epsilon^{abc}  \f{x^b}{r} \partial_i  \f{x^c}{r},\\
 g' Y_i
 &=& -\sin^2 \theta_W j(\theta) (1+\cos \theta )\partial_i \varphi.
\end{eqnarray}
with $\theta_W$ the Weinberg angle.
Here $f(\theta)$, $h(\theta)$ and $j(\theta)$ are profile functions vanishing on $\theta=0$ 
(the positive side of the $z$ axis):
\begin{equation}
 f(0)= h(0) = j(0)=0,
\end{equation}
and thus the electroweak symmetry is restored ($|\Phi_\mr{SM}| =0$) on $\theta=0$ corresponding to the $Z$ string (string-like defect).

As we move away from $\theta=0$ on the large sphere, the profile functions approach to unity 
and then we obtain the following asymptotic forms:
\begin{eqnarray}
 \Phi_{\mathrm{SM}} &=& v \begin{pmatrix}
  e^{ i\varphi} \cos \f{\theta}{2} \\  \sin \f{\theta}{2}
    \end{pmatrix},\label{041435_11Jun20}\\
 g W_{i} ^{a
 } &=& -\cos^2 \theta_W n^a (1+\cos \theta) \partial_i \varphi-\epsilon^{abc}  \f{x^b}{r} \partial_i  \f{x^c}{r} \label{042654_11Jun20},\\
 g' Y_i
 &=& -\sin^2 \theta_W (1+\cos \theta )\partial_i \varphi\label{042659_11Jun20}.
\end{eqnarray}
These configurations seem to have a line singularity on $\theta =0$ corresponding to the $Z$ string
and thus are valid only outside of the string.
Note that the $Z$ string is a regular and physical object unlike the Dirac string.

We introduce a unit vector as
\begin{equation}
  n^a_{\mathrm{SM}} \equiv \f{\Phi_{\mathrm{SM}}^\dagger \sigma^a \Phi_{\mathrm{SM}}}{|\Phi_{\mathrm{SM}}|^2},\label{204610_23Jun20}
\end{equation}
transforming as the adjoint representation of $SU(2)_W$, 
which is analogours to that of the 't Hooft-Polyakov monopole \eqref{181300_26Jun20}.
Substituting Eqs.~\eqref{041435_11Jun20}-\eqref{042659_11Jun20}, we have
\footnote{For the Higgs field to be regular at the center of the string, $\rho=0$, the function $f(\theta)$ satisfies $f(\theta)\sim \sin\theta$ for $\theta\to 0$ \cite{Vachaspati:1992fi,Nielsen:1973cs,Abrikosov:1956sx}.}:
\begin{align}
  n^a_{\mathrm{SM}} &= \frac{1}{f^2 c_{1/2}^2 + s_{1/2}^2 }
  \left(f\f{x}{r}, f\f{y}{r}, f^2 c_{1/2}^2 - s_{1/2}^2\right)\nonumber\\
&= \left(\sin \Theta_{\mathrm{SM}} \cos \varphi, \sin \Theta_{\mathrm{SM}} \sin \varphi,\cos\Theta_{\mathrm{SM}} \right)
\end{align}
where we have defined 
\begin{equation}
 s_{1/2} \equiv \sin \f{\theta}{2}, \h{2em}  c_{1/2} \equiv \cos \f{\theta}{2}
\end{equation}
and 
\begin{align}
  \sin \Theta_{\mathrm{SM}} &\equiv \frac{f}{f^2 c_{1/2}^2 + s_{1/2}^2 } \f{\sqrt{x^2+y^2}}{r}\\
 \cos \Theta_{\mathrm{SM}} &\equiv \frac{f^2 c_{1/2}^2 - s_{1/2}^2}{f^2 c_{1/2}^2 + s_{1/2}^2 } .
\end{align}
Note that $n^a_{\mathrm{SM}}$ is {\it singular} at $\theta=0$
because the electroweak gauge symmetry is restored at $\theta=0$ ($|\Phi_\mr{SM}| =0$).

 We can define a $U(1)$ subgroup of $SU(2)_W$ that keeps the vector $n^a_{\mathrm{SM}}$ invariant, which is denoted by $U(1)_n$.%
\footnote{Although $U(1)_n$ keeps $n^a_\mr{SM}$ invariant, it does not keep $\Phi_\mr{SM}$ invariant,
thus $U(1)_n$ itself is not an unbroken subgroup.
}
The electromagnetic and $Z$ field strengths are defined as superpositions of $U(1)_n$ and $U(1)_Y$ \cite{Nambu:1977ag}:
\begin{equation}
 F_{ij}^\mr{EM} \equiv -\sin \theta _W n^a_{\mathrm{SM}} W_{ij}^a  + \cos \theta_W  Y_{ij},\label{045908_11Jun20}
\end{equation}
\begin{equation}
 F_{ij}^{Z} \equiv -\cos \theta _W n^a_{\mathrm{SM}} W_{ij}^a  -\sin \theta_W  Y_{ij}\label{171932_26Jun20}.
\end{equation}
From the fact that 
we cannot define the subgroup $U(1)_n$ at $\theta=0$ because $n^a_{\mathrm{SM}}$ is ill-defined there,
it follows that the electromagnetic and $Z$ field strengths, \eqref{045908_11Jun20} and \eqref{171932_26Jun20}, are not defined at $\theta=0$.
This is an important difference from the 't Hooft-Polyakov case.

To see the electromagnetic and $Z$ fluxes at large distances from the string core, 
it is convenient to use the asymptotic forms in Eqs.~\eqref{041435_11Jun20}-\eqref{042659_11Jun20} at $\theta \gg 0$.
From Eq.~\eqref{041435_11Jun20}, we can see that $n^a_{\mathrm{SM}}$ approaches to a hedgehog structure,
\begin{equation}
 n^a_{\mathrm{SM}} = (\sin \theta \cos \varphi, \sin \theta \sin \varphi, \cos \theta)^t = \frac{x^a}{r}.
\end{equation}
Plugging this and the ansatz in Eq.~\eqref{041435_11Jun20}-\eqref{042659_11Jun20} 
into Eqs.~\eqref{045908_11Jun20} and \eqref{171932_26Jun20}, 
we get the physical field strengths at large distances as
\begin{align}
 F_{ij}^Z &=  \f{4\pi \cos \theta_\mr{W}}{g} \theta(z)  \epsilon_{3ij}\delta (x) \delta (y),\label{172051_26Jun20}\\
 F_{ij}^\mr{EM} &=  \f{\sin \theta_\mr{W}}{g} \epsilon^{aij} \f{x^a}{r^3}.\label{172031_26Jun20}
\end{align}
Note that the $\delta$-function singularity at $\theta=0$ in $F_{ij}^Z$ comes from  the asymptotic forms at $\theta \gg 0$ 
and the true form has a finite width (the $Z$ string is regular solution),  
decaying exponentially like $\sim \exp(-m_Z \rho)$.
From Eq.~\eqref{172031_26Jun20}, it is clear that there is a magnetic flux from the origin in a spherical hedgehog form.
The total amount of the magnetic flux $\Phi_{\rm EM}$ can be calculated by integrating the flux density 
$B_i \equiv \f{1}{2}\epsilon_{ijk}F_{jk}^\mr{EM}$ as
\begin{equation}
 \Phi_{\rm EM} = \int d^3x~ \partial_i B_i =\f{4 \pi \sin \theta_\mr{W}}{g}.
\end{equation}
In addition, from Eq.~\eqref{172051_26Jun20},
the $Z$-fluxes only exist on the positive side of the $z$-axis as
\begin{eqnarray}
\Phi_Z\big|_{z > 0} &=& \int {d^2x} \, F_{ij}^Z\big|_{z > 0} = \frac{4\pi\cos\theta_\mr{W}}{g},\label{204225_3Jul20}
\end{eqnarray}
which flows on the $z$ axis from the origin.
Therefore, the former is indeed yeilding a long-range force 
while the latter is massive and confined around the string on $\theta=0$.
See Fig.~\ref{040735_1May20} for a schematic picture of the $Z$- and  electromagnetic fluxes for the Nambu monopole in the SM.

\subsection{Fiber bundle}
 \begin{figure}[tbp]
 \centering
 \includegraphics[width=0.5\textwidth]{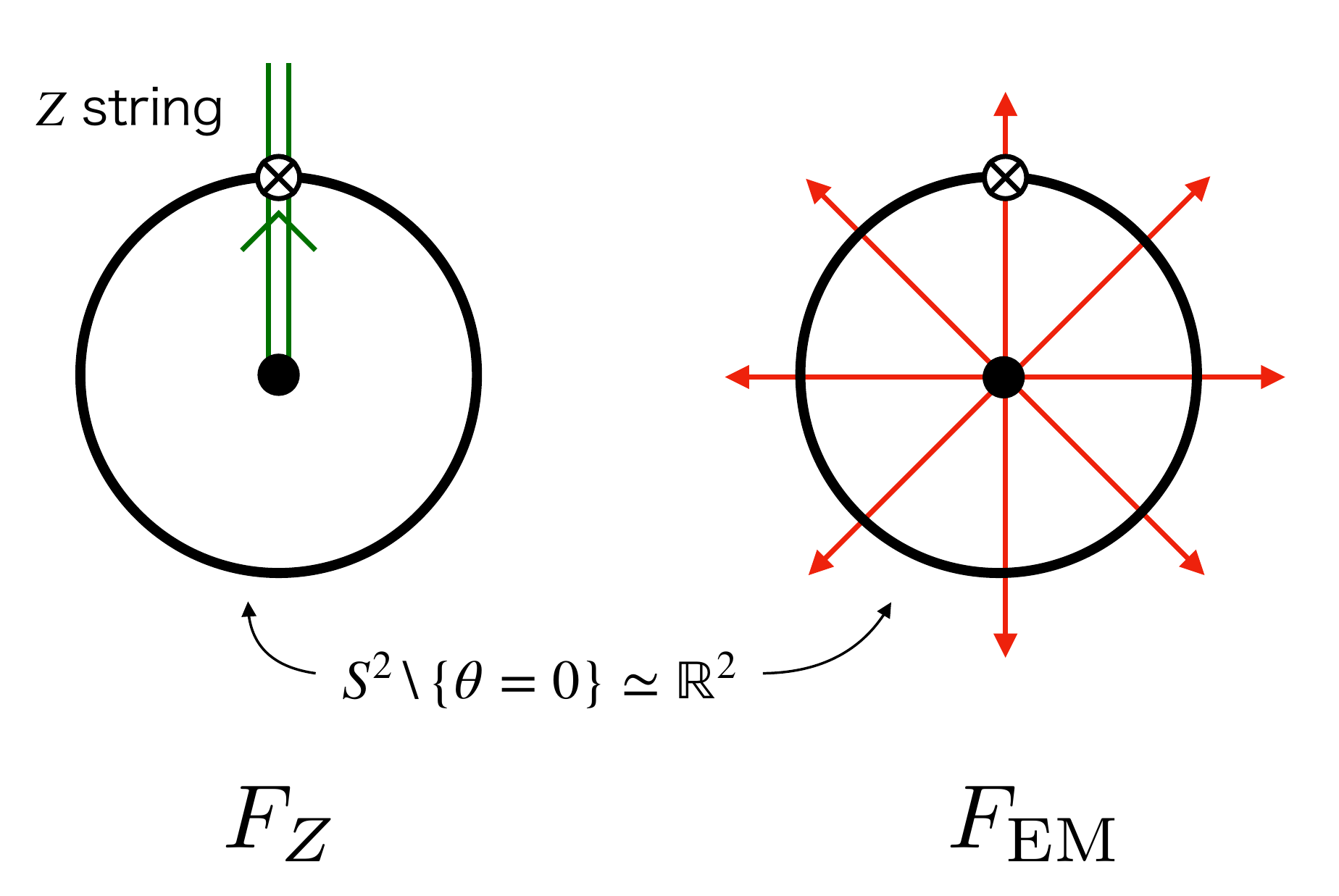}
  \caption{
  Schematic picture of $S^2$ surrounding the Nambu monopole in the SM.
  The $Z$ string passes through the north pole on $S^2$,
  and makes the electromagnetic gauge field singular at the string core $\theta =0$.
  As a result, it is well-defined only on $S^2\backslash \{\theta =0 \} \simeq \mathbb{R}^2$.
  Such a base space can be spanned by a single patch.
  }
  \label{040735_1May20}
 \end{figure}

To clarify a (non-)topological property of the Nambu monopole in the SM, 
let us discuss the fiber bundle of the electromagnetic gauge group $U(1)_{\rm EM}$.
We consider a two-dimensional sphere $S^2$ surrounding the monopole at spatial infinity,
where the single $Z$-string passes through the sphere (See Fig.~\ref{040735_1May20}).
Here, we have an important remark,
that is, the vector $n^a_\mr{SM}$ indicating the subgroup $U(1)_n$ of $SU(2)_W$ is ill-defined at the string center $\theta=0$.
As a result, we cannot define $U(1)_n$ and hence $U(1)_{\rm EM}$ at the north pole $\theta=0$ on $S^2$.
Thus we must consider the base space $S^2 \backslash \{\theta=0\} \simeq \mathbb{R}^2$ with the fiber $U(1)_{\rm EM}$, instead of $S^2$.
Obviously, we can span $\mathbb{R}^2$ by a single patch and the fiber bundle is trivial: $\mathbb{R}^2 \times U(1)_{\rm EM} (\simeq \mathbb{R}^2 \times S^1)$.
Therefore, this configuration is topologically distinct from the previous case of the 't Hooft-Polyakov monopole, and its topological structure is trivial.
As a result of the fact that the monopole has no topological origin, 
{\it i.e.}, the trivial fiber bundle: $\mathbb{R}^2 \times U(1)_{\rm EM}$,
the monopole does not provide the Dirac's quantization condition,
and thus the magnetic charge is not quantized.
Indeed, the monopole can be removed by a continuous deformation;
for instance, we can deform the $Z$ string into 
the vacuum since the $Z$ string is not topologically protected 
($\pi_1(S^3)=0$ for the vacuum manifold $S^3$ of the SM), and hence the monopole disappears.

\section{Definition of electromagnetic field in 2HDM}
\label{044315_30Jul20}
\subsection{Two Higgs doublet model}
Before turning into the Nambu monopole in 2HDM, we here clarify the definition of the electromagnetic gauge group in 2HDM.
In a homogeneous vacuum in the absence of any solitons, we can define the electromagnetic field and the field strength obviously 
by a projection of the gauge fields onto the unbroken direction.
On the other hand, it is non-trivial in the presence of solitons, 
in which the Higgs fields, {\it i.e.}, the unbroken direction of the gauge symmetry, varies in the space.
The definition without solitons should be replaced by a more general definition for the cases with varying VEVs.
In the SM, the electromagnetic field strength is defined by Eq.~\eqref{045908_11Jun20}, which is given by Nambu in Ref.~\cite{Nambu:1977ag}.
There are also other definitions, e.g., those in Refs.~\cite{Hindmarsh:1993aw,Tornkvist:1998sm},
although they are equivalent at large distances from the solitons, in which $D_i \Phi_\mr{SM}\sim 0$ and $|\Phi_\mr{EM}|\sim v$.
In 2HDM, on the other hand, {\it general definitions} for the electromagnetic field that can be apply to the varying VEVs 
has not been given in the literature.
For later use, we present such a {\it general definition} of the electromagnetic field in the presence of solitons in 2HDM for the first time.

We start with the Lagrangian of the electroweak sector in 2HDM,
in which there are two Higgs doublets, $\Phi_1$ and $\Phi_2$ with the same $U(1)_Y $ hyper charge $+1$.
The Lagrangian is given by
\begin{align}
 \mathcal{L} &= -\f{1}{4}W_{\mu\nu}^a W ^{a\mu\nu} - \f{1}{4} Y_{\mu\nu} Y^{\mu\nu} + \mathrm{Tr} | D_\mu H |^2 - V(H),
\end{align}
where $W_\mu$ and $Y_\mu$ are the $SU(2)_W$ and $U(1)_Y$ gauge fields, respectively. 
We have adopted the $2\times 2$ matrix notation \cite{Grzadkowski:2010dj}: $H \equiv (i\sigma_2 \Phi_1^\ast, \Phi_2)$.
The matrix field $H$ transforms under the electroweak $SU(2)_W \times U(1)_Y$ symmetry as
\begin{equation}
H \to \exp\left[\f{i}{2}\theta_a(x) \sigma_a\right] H ~\exp\left[-\f{i}{2} \theta_Y(x) \sigma_3\right],
\end{equation}
where the group element acting from the left belongs to $SU(2)_W$ and the other element acting from the right belongs to $U(1)_Y$.
Therefore the covariant derivative on $H$ can be expressed as
\begin{equation}
D_\mu H =\partial_\mu H - i \frac{g}{2} \sigma_a W_\mu^a H + i \frac{g'}{2}H\sigma_3 Y_\mu.
\end{equation}
The VEVs of $H$ is expressed by a diagonal matrix $\langle H \rangle = \mr{diag} (v_1,v_2)$,
and the Higgs potential can be written by using $H$ as follows:
\begin{align}
V(H)
& = - m_{1}^2~ \mr{Tr}|H|^2 - m_{2}^2~ \mr{Tr}\left(|H|^2 \sigma_3\right) -  \left( m_{3}^2 \det H  + {\rm h.c.} \right)\nonumber\label{202504_23Jun20} \\
& + \alpha_1~\mr{Tr}|H|^4  +  \alpha_2 ~\left(\mr{Tr}|H|^2 \right)^2+ \alpha_3~ \mr{ Tr}\left(|H|^2 \sigma_3 |H|^2\sigma_3\right)  \n \\
& + \alpha_4~ \mr{Tr}\left(|H|^2 \sigma_3 |H|^2\right) + \left( \alpha_5 ~\det H^2 + {\rm h.c.}\right),
\end{align}
where $|H|^2 \equiv H ^\dagger H$ and we have imposed a (softly-broken) $\mathbb{Z}_2$ symmetry: $\Phi_1 \to +\Phi_1, \Phi_2 \to -\Phi_2$ (or equivalently, $H \to H \sigma_3$).
The relations between the parameters in Eq.~\eqref{202504_23Jun20} and the conventional parameterization are shown in Ref.~\cite{Eto:2020hjb}.

In the next section, we will consider the topologically stable Nambu monopole in 2HDM.
The stability is realized by setting $m_3 = \alpha_4=\alpha_5=0$ and $\alpha_3 <0$ in the potential \cite{Eto:2020hjb},
resulting in two global symmetries:
\begin{equation}
 \begin{cases}
  U(1)_a :& H \to e^{i \alpha}H \\
  (\mathbb{Z}_2)_\mr{C} :& H \to (i\sigma_2)^\dagger H (i\sigma_2)
  ,\quad W_\mu \to (i\sigma_2)^\dagger W_\mu (i\sigma_2),\quad Y_\mu \to - Y_\mu\,.
 \end{cases}\label{200609_24Jun20}
\end{equation}
As a result of the $(\mathbb{Z}_2)_C$ symmetry, we have $v_1 = v_2$ ($\tan \beta \equiv v_2 / v_1 =1$).%
\footnote{The reader should not confuse the $(\mathbb{Z}_2)_\mr{C}$ symmetry with the softly-broken $\mathbb{Z}_2$ symmetry: $\Phi_1 \to +\Phi_1, \Phi_2 \to -\Phi_2$. }

Using the two doublets, we define two unit vectors 
\begin{equation}
 n_1^a \equiv \f{\Phi_1^\dagger \sigma^a \Phi_1}{|\Phi_1|^2}, \quad
 n_2^a \equiv \f{\Phi_2^\dagger \sigma^a \Phi_2}{|\Phi_2|^2},\label{080112_1Apr20}
\end{equation}
which are analogous to Eq.~\eqref{204610_23Jun20} in the SM
and are ill-defined when $|\Phi_1|=0$ and $|\Phi_2|=0$, respectively.

\subsection{$n_1^a=n_2^a$ case}
We first assume that $n_1^a = n_2 ^a$ holds in regions where $\Phi_1\neq 0$ and $\Phi_2 \neq 0$,
which we will encounter in the subsequent sections.
Taking the unitary gauge, this situation is realized when the doublets have the following forms:
\begin{equation}
 \Phi_1=\begin{pmatrix}
0 \\ \Phi_{1,2}
	\end{pmatrix},
\h{2em}
 \Phi_2= \begin{pmatrix}
0 \\ \Phi_{2,2}	 
	\end{pmatrix},
\end{equation}
where $\Phi_{1,2}$ and $\Phi_{2,2}$ are complex functions.
This configuration is sometimes called the neutral configuration in the literature.
We will relax this condition later.

In the presence of solitons, the Higgs fields $H$ acquire $x$-dependent VEVs,
and then the gauge fields obtain $x$-dependent masses as follows:
\begin{align}
\mathrm{Tr}|D _\mu H|^2 &= |D_\mu \Phi_1|^2 +|D_\mu \Phi_2|^2 \nonumber \\
 &\supset \sum_{f=1,2} \left|\left(-\f{i}{2}g W_\mu^a \sigma^a - \f{i}{2}g' Y_\mu \right) \Phi_f \right|^2 \nonumber \\
 &=\f{1}{4} \sum_f \Phi_f ^\dagger \left( g^2 W_\mu ^a W^{b\mu} \sigma^a \sigma^b + g'^2 Y_\mu Y^\mu + 2 g g' W_\mu^a Y^\mu \sigma ^a\right) \Phi_f  \nonumber \\
 &=\f{1}{4} \sum_f |\Phi_f |^2 \left( g^2 W_\mu ^a W^{a\mu} + g'^2 Y_\mu Y^\mu\right)  
+ 2 g g' \f{1}{4} \sum_f |\Phi_f |^2  n_f^a  W_\mu^a Y^\mu.\label{063009_1Apr20}
\end{align}
In order to rewrite Eq.~\eqref{063009_1Apr20},
we introduce an adjoint unit vector $n^a$ as
\begin{align}
 &\sum_f |\Phi_f |^2  n_f^a \equiv n^a  \sum_f |\Phi_f |^2 
 \nonumber 
 \\
 \Leftrightarrow~& n^a \equiv \f{\Phi_1^\dagger\sigma^a \Phi_1 + \Phi_2^\dagger\sigma^a \Phi_2} {|\Phi_1 |^2 +|\Phi_2 |^2} = \f{\sum_f|\Phi_f |^2\, n^a_f} {\sum_f|\Phi_f |^2} ,\label{140145_1Apr20}
\end{align}
which is well-defined everywhere except for $|\Phi_1| = |\Phi_2|=0$.
We thus have
\begin{align}
  \eqref{063009_1Apr20}
 &=\f{1}{4} \sum_f |\Phi_f |^2 \left( g  n^a W_\mu ^a + g' Y_\mu \right)^2
 +  \f{g^2}{4} \sum_f |\Phi_f |^2 \left( \delta^{ab} - n^a n^b \right) W_\mu ^a W_\mu ^b \nonumber \\
 &\equiv \f{g_Z^2}{4} \sum_f |\Phi_f |^2 Z_\mu Z^\mu 
  + \f{g^2}{4} \sum_f |\Phi_f |^2 ~\mathrm{Tr} (W_\mu^\perp W^{\perp \mu} )
\end{align}
where we have defined 
\begin{eqnarray}
 Z_\mu &\equiv& - \cos\theta_W n^a W_\mu ^a - \sin\theta_W Y_\mu,\label{065506_1Apr20}\\
 W_\mu ^\perp &\equiv& \f{\sigma^a}{2} \left(\delta_{ab} - n_a n_b \right) W_\mu^b,\label{065511_1Apr20}
\end{eqnarray}
and $g_Z \equiv \sqrt{g^2 + g'{}^2}$.
Eqs.~\eqref{065506_1Apr20} and \eqref{065511_1Apr20} are the neutral $Z$ boson and the charged $W$ boson, respectively.%
\footnote{Note that $W_\mu^\perp$ itself is not an eigenstate of $U(1)_{\rm EM}$.}
On the other hand, the massless electromagnetic gauge field (photon) is defined by
\begin{equation}
 A_\mu \equiv - \sin\theta_W n^a W_\mu ^a + \cos\theta_W Y_\mu,\label{005733_24Jun20}
\end{equation}
which is orthogonal to the $Z$ field.
We stress that the electromagnetic and the massive gauge bosons can be defined everywhere unless $\Phi_1 = \Phi_2=0$.

Similarly, we can define the electromagnetic and $Z$ field strength using $n^a$ as
\begin{equation}
 F_{\mu\nu}^{Z} \equiv  - \cos\theta_W n^a W_{\mu\nu} ^a - \sin\theta_W Y_{\mu\nu},\label{231040_24Jun20}
\end{equation}
\begin{equation}
 F_{\mu\nu}^\mr{EM} \equiv  - \sin\theta_W n^a W_{\mu\nu} ^a + \cos\theta_W Y_{\mu\nu},\label{004333_1May20}
\end{equation}
respectively. 
Indeed, the latter satisfies the source-free Maxwell equations
\begin{equation}
\partial^\mu F_{\mu\nu}^\mr{EM}=0, \h{2em} \partial^\mu \tilde{F}_{\mu\nu}^\mr{EM}=0
\end{equation}
at large distances, and yields the long-range force.
Note that these definitions are analogous to those in the SM, but $n_\mr{SM}^a$ is replaced by $n^a$ \eqref{140145_1Apr20}.

\subsection{$n_1^a \neq n_2^a$ case}
We consider the case of $n_1^a \neq n_2^a$.
In the unitary gauge, this is realized when, e.g.,
\begin{equation}
 \Phi_1=\begin{pmatrix}
0 \\ \Phi_{1,2}
	\end{pmatrix},
\h{2em}
 \Phi_2= \begin{pmatrix}
\Phi_{2,1} \\ \Phi_{2,2}	 
	\end{pmatrix},
\end{equation}
where $\Phi_{1,2}$, $\Phi_{2,1}$ and $\Phi_{2,2}$ are complex functions.
In the literature, this is sometimes called the {\it charged} configuration.
In such a case, there is no unbroken subgroup in the $SU(2)_W \times U(1)_Y$ symmetry, 
thus $U(1)_\mr{EM}$ is also broken.
To see this, we start with the mass terms for the gauge fields again:
\begin{align}
\mathrm{Tr}|D _\mu H|^2 
 &\supset \sum_{f=1,2} \left|\left(-\f{i}{2}g W_\mu^a \sigma^a - \f{i}{2}g' Y_\mu \right) \Phi_f \right|^2 \nonumber \\
 &=\f{1}{4} \sum_f |\Phi_f |^2 \left( g^2 W_\mu ^a W^{a\mu} + g'^2 Y_\mu Y^\mu\right) 
 + 2 g g' \f{1}{4} \sum_f |\Phi_f |^2  n_f^a  W_\mu^a Y^\mu.\label{144719_1Apr20}
\end{align}

Let us rewrite the last term in Eq.~\eqref{144719_1Apr20}.
Note that the vector $n^a$ defined in Eq.~\eqref{140145_1Apr20} is no longer normalized to unity because of $n^a_1 \neq n^a_2$.
Thus, we introduce an alternative unit vector $\tilde{n}^a$ as
\begin{align}
 &\sum_f |\Phi_f |^2  n_f^a \equiv C \tilde{n}^a \nonumber \\
\Leftrightarrow~& \tilde{n}^a \equiv \f{\Phi_1^\dagger\sigma^a \Phi_1 + \Phi_2^\dagger\sigma^a \Phi_2} {C},\label{215346_27May20}
\end{align}
where $C$ is a normalization factor:
\begin{align}
  C^2 &= \left(\Phi_1^\dagger\sigma^a \Phi_1 + \Phi_2^\dagger\sigma^a \Phi_2 \right)^2, \label{011907_24Jun20}
\end{align}
and is taken as $C>0$.

Using Eq.~\eqref{215346_27May20} and dividing $(W_\mu^a)^2$ into two parts, 
we can rewrite Eq.~\eqref{144719_1Apr20} as
\begin{align}
\eqref{144719_1Apr20}  
%
& =\f{1}{4} \sum_f |\Phi_f |^2  g^2 \left( \tilde{n}^a W_\mu ^a \right)^2 + \f{1}{4} \sum_f |\Phi_f |^2 g'^2 \left( Y_\mu \right)^2   \n \\
 &\h{3em} +2 g g' \f{1}{4} C  \tilde{n}^a  W_\mu^a Y^\mu 
+ \f{g^2}{4} \sum_f |\Phi_f |^2 \left(\delta^{ab}- \tilde{n}^a \tilde{n}^b \right) W_\mu^a  W^{b\mu} \nonumber \\
&= \f{1}{4 } 
 \begin{pmatrix}
  \tilde{n}^a W_\mu ^a \\   Y_\mu
 \end{pmatrix}^T
 M
 \begin{pmatrix}
  \tilde{n}^a W_\mu ^a \\  Y_\mu
 \end{pmatrix} 
+ \f{g^2}{4} \sum_f |\Phi_f |^2~ \mathrm{Tr} ( W_\mu^\perp  W^{\perp \mu}) 
\end{align}
with
\begin{equation}
 M \equiv  \begin{pmatrix}
 g^2 \sum_f |\Phi_f |^2 & g g' C \\ g g' C &  g'^2\sum_f |\Phi_f |^2
 \end{pmatrix}\label{152726_1Apr20}
\end{equation}
and 
\begin{equation}
 W ^\perp _\mu \equiv \f{\sigma^a}{2} \left(\delta^{ab}- \tilde{n}^a \tilde{n}^b \right) W_\mu^b,
\end{equation}
where $ W ^\perp _\mu $ is the orthogonal component to $\tilde{n}^a$, similarly to Eq.~\eqref{065511_1Apr20}.

The mass matrix $M$ in \eqref{152726_1Apr20} provides masses for the $Z$ gauge boson and photon.
Unlike the previous case, the matrix have two non-zero eigenvalues, 
and thus there is no massless field.
However, we still can define a photon as the lighter component among the two massive gauge bosons.
To achieve this, let us diagonalize the mass matrix $M$ by the following basis transformation:
\begin{align}
 \begin{pmatrix}
  \tilde{n}^a W_\mu ^a \\   Y_\mu
 \end{pmatrix}
\to 
U \begin{pmatrix}
  \tilde{n}^a W_\mu ^a \\   Y_\mu
 \end{pmatrix},
\end{align}
with 
\begin{equation}
 U \equiv \begin{pmatrix}
  \cos \xi & \sin \xi \\  -\sin \xi & \cos \xi
 \end{pmatrix},
\end{equation}
where $\xi$ is an ``effective Weinberg angle'' satisfying
\begin{equation}
 \tan 2\xi = r_C \tan 2 \theta_W
\end{equation}
with $r_C \equiv \sum_f |\Phi_f|^2 /C$.

The mass eigenvalues and the mass eigenstates are 
\begin{align}
\f{1}{4} \begin{pmatrix}
  \tilde{n}^a W_\mu ^a \\   Y_\mu
 \end{pmatrix}^T
 M
 \begin{pmatrix}
  \tilde{n}^a W_\mu ^a \\  Y_\mu
 \end{pmatrix}  
%
&= \f{m_Z^2}{2} Z_\mu Z^\mu + \f{m_\gamma^2}{2} A_\mu A^\mu
\end{align}
with
\begin{equation}
 m_Z^2 \equiv \f{\sum _f |\Phi_f|^2 }{4} \left( 1+\cos 2\theta_W \cos 2\xi + r_C \sin 2\theta_W \sin 2 \xi \right)\label{014011_4Apr20}
\end{equation}
\begin{equation}
 m_\gamma^2 \equiv \f{\sum _f |\Phi_f|^2 }{4} \left( 1 - \cos 2\theta_W \cos 2\xi - r_C \sin 2\theta_W \sin 2 \xi \right).\label{011635_24Jun20}
\end{equation}
and 
\begin{equation}
 Z_\mu \equiv   - \cos\xi \, \tilde{n}^a W_\mu ^a - \sin\xi \, Y_\mu,\label{013338_24Apr20}
\end{equation}
\begin{equation}
 A_\mu \equiv   - \sin\xi \, \tilde{n}^a W_\mu ^a + \cos\xi \, Y_\mu.\label{014017_4Apr20}
\end{equation}
Note that $A_\mu$ is the electromagnetic gauge field (photon), but it is no longer massless, $m_\gamma \neq 0$.
For arbitrary $\theta_W$ and $C\neq 0$, $m_\gamma^2 < m_Z^2$ always holds.

To compare with the previous result in the last subsection,
let us reproduce the case of $n_1^a=n_2^a$.
It leads to $n^a=\tilde{n}^a$,
and hence $r_C=1$ and $\xi=\theta_W$.
The mass eigenstates, Eqs.~\eqref{013338_24Apr20} and \eqref{014017_4Apr20}, reduce to the previous result, Eqs.~\eqref{065506_1Apr20} and \eqref{005733_24Jun20}.
The mass eigenvalues \eqref{014011_4Apr20} and \eqref{011635_24Jun20} reduce to
\begin{equation}
 m_Z^2 \to \f{\sum _f |\Phi_f|^2 }{2}, \h{2em} m_\gamma^2 \to 0,
\end{equation}
which agree with the previous ones.

Note that we cannot use the above definition for $C=0$, in which the vector $\tilde{n}^a$ is not well-defined.
Before closing this section, let us see when such a case occurs.
Eq.~\eqref{011907_24Jun20} can be rewritten as
\begin{align}
  C^2 
&= \left(\Phi_1^\dagger \Phi_1\right)^2 + \left(\Phi_2^\dagger \Phi_2\right)^2 + 2 \left(\Phi_1^\dagger\sigma^a \Phi_1\right) \left( \Phi_2^\dagger\sigma^a \Phi_2\right) .
\end{align}
Using the Fierz identity \eqref{142651_1Apr20}, we have
\begin{align}
 2 \left(\Phi_1^\dagger\sigma^a \Phi_1\right) \left( \Phi_2^\dagger\sigma^a \Phi_2\right) 
&= 4 | \Phi_1^\dagger \Phi_2|^2 - 2 |\Phi_1|^2 |\Phi_2|^2,
\end{align}
and thus 
\begin{align}
 C^2 &= (|\Phi_1|^2 - |\Phi_2|^2)^2 + 4 | \Phi_1^\dagger \Phi_2|^2 \geq 0.\label{215618_27May20} 
\end{align}
The equality holds when 
\begin{equation}
 |\Phi_1|^2 = |\Phi_2|^2, \h{2em} \Phi_1^\dagger \Phi_2=0,
\end{equation}
in which the second term in Eq.~\eqref{144719_1Apr20} vanishes,
and thus all $U(1)$ subgroups in $SU(2)_W$ are equivalent and not mixed with the $U(1)_Y$ component.

\section{Topological property of Nambu monopole in 2HDM}
\label{044348_30Jul20}
We here show that the Nambu monopole in 2HDM has a non-trivial topology as for the 't Hooft-Poyakov monopole and the Wu-Yang monopole,
unlike in the SM.
This is one of the main results of this paper.
In the first subsection, we first give a brief review on the electroweak strings and Nambu monopole in 2HDM.
The Nambu monopole in 2HDM is a magnetic monopole attached with {\it two topological Z strings} on the opposite sides 
\cite{Eto:2019hhf,Eto:2020hjb}.
After that, we see the topological structure, especially the fiber bundle of the electromagnetic gauge field of the monopole.
As we stated above, we assume the $U(1)_a$ and $(\mathbb{Z}_2)_\mr{C}$ symmetries \eqref{200609_24Jun20},
but they are not essential for the topolgoical structure of the fiber bundle.

\subsection{Nambu monopole in 2HDM}
We first consider vortex solutions in 2HDM.
As studied in Refs.~\cite{Dvali:1993sg,Dvali:1994qf,Battye:2011jj,Eto:2018tnk,Eto:2018hhg}, 
the 2HDM with the $U(1)_a$ symmetry, Eq.~\eqref{200609_24Jun20}, admits topological vortex solutions.
They can have (approximate) non-Abelian moduli (it can be genuine non-Abelian moduli in the limit of $g' \to 0$)
and confine the electroweak magnetic fluxes inside them, and hence are called the electroweak strings.

One of them is called the $(1,0)$-string,
whose field configuration is given by
\begin{align}
 H^{(1,0)} &= v\begin{pmatrix}
	         f^{(1,0)}(\rho) e^{i\varphi} &0\\
               0 & h^{(1,0)}(\rho)
	      \end{pmatrix},\label{eq:H(1,0)}\\
 Z_i ^{(1,0)} &=   - \f{\cos \theta_\mr{W}}{g} \f{ \epsilon_{3ij}x^j}{\rho^2} (1-w^{(1,0)}(\rho)) \label{013304_3Dec19},
\end{align}
where $\rho \equiv \sqrt{x^2+y^2}$ and $\varphi$ is the rotation angle around the $z$-axis.
The functions $f^{(1,0)}$, $h^{(1,0)}$ and $w^{(1,0)}$ satisfy the following boundary conditions
\begin{equation}
 f^{(1,0)}(\infty)= h^{(1,0)}(\infty)=1,\h{1em}w^{(1,0)}(\infty)=0,\label{181958_12Jul20}
\end{equation}
\begin{equation}
 f^{(1,0)}(0)= \partial_\rho h^{(1,0)}(0)=0, \h{1em} w^{(1,0)}(0)=1 ,\label{182002_12Jul20}
\end{equation}
so that the configurations are regular at $\rho=0$.
The precise forms of the functions are determined by solving the EOMs, 
but they are irrelevant in our following argument.
In Eq.~\eqref{013304_3Dec19}, we have used the definition of the $Z$ gauge field, Eq.~\eqref{065506_1Apr20}, 
but it becomes trivial since the unit vector $n^a$ is constant $n^a=(0,0,-1)$ in the vortex configuration Eq.~\eqref{eq:H(1,0)}.

Due to the $(\mathbb{Z}_2)_\mr{C}$ symmetry in Eq.~\eqref{200609_24Jun20}, 
we have another stable vortex solution called the $ (0,1) $-string:
\begin{align}
 H^{(0,1)} &=  v \begin{pmatrix}
	        h^{(0,1)}(\rho) &0 \\
               0 & f^{(0,1)}(\rho) e^{i\varphi} 
	      \end{pmatrix},\label{182742_12Jul20}\\
 Z_i ^{(0,1)} &= \f{  \cos \theta_\mr{W}}{g}  \f{ \epsilon_{3ij}x^j}{\rho^2} (1-w^{(0,1)}(\rho))\label{012222_17Mar19},
\end{align}
where the profile functions $f^{(0,1)}$, $h^{(0,1)}$ and $w^{(0,1)}$ satisfy the same boundary conditions with Eqs.~\eqref{181958_12Jul20} and \eqref{182002_12Jul20}.
The $(0,1)$-string has the degenerated tension with that of the $(1,0)$-string as a result of the $(\mathbb{Z}_2)_\mr{C}$ symmetry.%
\footnote{One can check that they are related by the $(\mathbb{Z}_2)_\mr{C}$ transformation \eqref{200609_24Jun20}.}
These configurations are not in the vacuum around $\rho=0$, 
{\it i.e.}, the cores of the two $Z$-strings.
Note that, in contrast to the case of the SM, {\it the electroweak symmetry is not restored} even around the string cores $\rho\sim 0$
because one doublet vanishes on $\rho \sim 0$ while the other does not (for instance, $f^{(1,0)}(0)= 0$ but $h^{(1,0)}(0)\neq 0$).
Therefore, one can define the electromagnetic field even inside the strings. This fact plays a crucial role for the topological structure of the Nambu monopole in 2HDM as seen below.

By rewriting Eqs.~\eqref{eq:H(1,0)} and \eqref{182742_12Jul20} as
\begin{equation}
H^{(1,0)}= v e^{i\f{\varphi}{2} } e^{i\f{\varphi}{2} \sigma_3} \begin{pmatrix}
	         f^{(1,0)} &0\\
               0 & h^{(1,0)}
	      \end{pmatrix},
\end{equation}
\begin{equation}
H^{(0,1)}= v e^{i\f{\varphi}{2} } e^{-i\f{\varphi}{2} \sigma_3} \begin{pmatrix}
	         h^{(0,1)} &0\\
               0 & f^{(0,1)}
	      \end{pmatrix},
\end{equation}
it is clear that 
both the $(1,0)$- and $(0,1)$-strings have winding number $1/2$ for the global $U(1)_a$ symmetry,
and thus they are topological vortex strings of the global type. 
Similarly to standard global vortices, their tensions (masses per unit length) logarithmically diverge. 

On the other hand,
they also have a winding number $\pm 1/2$ inside the gauge orbit $U(1)_Z\in SU(2)_W\times U(1)_Y$,
which lead to the $Z$ fluxes flowing inside them.
The amounts of the fluxes of $(1,0)$- and $(0,1)$-string are
\begin{equation}
 \Phi_Z^{(1,0)}= \frac{2 \pi \cos \theta_\mr{W} }{ g}, \h{2em}  \Phi_Z^{(0,1)}= - \frac{2 \pi  \cos \theta_\mr{W} }{ g}, 
\label{eq:Z_fluxes}
\end{equation}
along the $z$-axis, respectively.
They are half of that of a non-topological $Z$ string in the SM because of the half winding number.
The $Z$ flux is squeezed into a flux tube.
In other words, contributions to the energy from the non-Abelian parts do not diverge.

There are physically different string configurations with a different (non-Abelian) magnetic flux, 
having a common winding number $1/2$ for the global $U(1)_a$.
Since they belong to the same topological sector classified by the first homotopy group $\pi_1(U(1)_a)$, 
one can deform one to the other continuously 
(with some energy cost).\footnote{
In a certain limit of 
$e=0$ with the custodial symmetry, they can continuously deformed to each other 
without energy cost by changing moduli.
}
Among them, the above two $Z$ strings are the most lightest configurations as long as $m_2 = \alpha_4 =0$ and $\alpha_3 \leq  0$ in the potential \eqref{202504_23Jun20} \cite{Eto:2018tnk,Eto:2020hjb},
and thus are energetically stable.

A configuration describing the Nambu monopole is obtained by continuously connecting the two $Z$ strings 
keeping the $U(1)_a$ topological winding number,
in which a junction point is a source for the $Z$ and the electromagnetic fluxes, and hence is a magnetic monopole.
We consider the case that $(1,0)$- and $(0,1)$-strings are put on the positive and negative side of the $z$-axis ($\theta=0,\pi$), respectively, 
and the monopole (junction point) is located at the origin.
At large distances from the origin, {\it i.~e.~}, $r\to \infty$, 
the Higgs field $H$ and the gauge fields $W_\mu,Y_\mu$ describing the monopole (with the strings) are given as
\begin{equation}
 H= v \begin{pmatrix}
     h_1(\theta) \sin \f{\theta}{2} & h_2(\theta)\cos \f{\theta}{2} \\
     - e^{i\varphi} f_1(\theta) \cos \f{\theta}{2} & e^{ i\varphi} f_2(\theta) \sin \f{\theta}{2}
    \end{pmatrix}\label{004019_25Jun20}
\end{equation}
\begin{equation}
 g W_i ^a = -\cos^2 \theta_W n^a j(\theta)\cos \theta \partial_i \varphi -\epsilon^{abc} n^b \partial_i n^c
\end{equation}
\begin{equation}
 g' Y_i = -\sin^2 \theta_W k(\theta)\cos \theta \partial_i \varphi.\label{195520_18Jul20}
\end{equation}
The profile functions $f_1,h_1$ and $f_2, h_2$ satisfy similar boundary conditions to those of the $(1,0)$- and $(0,1)$-strings on the north pole $\theta=0$ and the south pole $\theta=\pi$, respectively:
\begin{equation}
 f_1(0)=0,\h{2em} \partial_\theta h_1(0)=0,\label{014704_19Jul20}
\end{equation}
\begin{equation}
 f_2(\pi)=0, \h{2em} \partial_\theta h_2(\pi)=0,
\end{equation}
and they rapidly approach to unity as we move from the north and the south pole, respectively.
In addition, the profile functions $j$ and $k$ satisfy the following conditions:
\begin{equation}
 j(0)=j(\pi)= k(0)=k(\theta)=0\label{014708_19Jul20}
\end{equation}
with approaching to unity far away from the poles.
Eq.~\eqref{004019_25Jun20} indeed describes the $(1,0)$-string ($(0,1)$-string) on $\theta\sim 0$ ($\theta\sim\pi$) 
up to the $SU(2)_W$ gauge transformation.\footnote{
This is clear when one acts the $SU(2)_W$ gauge transformation $H\to UH$ with $U$ satisfying
\begin{equation*}
 U|_{\theta=\pi}= 1_{2\times2} , \h{2em}U|_{\theta=0}= -i \sigma_2 .
\end{equation*}
}
Thanks to the boundary conditions, the configuration does not have line singularities on $\theta=0,\pi$.
Note that one needs to introduce further $r$-dependent profile functions 
in order to smear the point-like singularity on $r=0$ and obtain a true solution of the EOMs.
Nevertheless, we do not care the singularity on $r=0$
because we are interested in topological properties of the monopole at large distances from the monopole $r\to \infty$.

It is worthwhile to see asymptotic forms of Eq.~\eqref{004019_25Jun20}-\eqref{195520_18Jul20} 
far away from the north and south poles.
This is equivalent to considering the infinitely thin monopole and strings.
Using the asymptotic behaviors in Eqs.~\eqref{014704_19Jul20}-\eqref{014708_19Jul20}, we have
\begin{equation}
 H= v \begin{pmatrix}
     \sin \f{\theta}{2} & \cos \f{\theta}{2} \\
     - e^{i\varphi} \cos \f{\theta}{2} & e^{ i\varphi} \sin \f{\theta}{2}
    \end{pmatrix}\label{200928_27May20}
\end{equation}
\begin{equation}
 g W_i ^a = -\cos^2 \theta_W n^a \cos \theta \partial_i \varphi -\epsilon^{abc} n^b \partial_i n^c\label{201019_27May20}
\end{equation}
\begin{equation}
 g' Y_i = -\sin^2 \theta_W \cos \theta \partial_i \varphi.\label{201042_27May20}
\end{equation}
This asymptotic configuration seems to have line singularities on $\theta=0$ and $\pi$, 
which correspond to the cores of the two $Z$-strings; $(1,0)$- and $(0,1)$-strings, respectively, as artifacts of asymptotic forms, 
and is valid only at large distances from the line singularities, $\rho\to \infty$ 
(Indeed, the singularities are avoided because $f_1(0) = 0$ and $f_2(\pi)=0$ in Eq.~(\ref{004019_25Jun20}).

Let us see the $Z$ and the electromagnetic fluxes around the monopole.
To do so, we first consider the vectors $n_1^a$ and $n_2^a$ defined by Eq.~\eqref{080112_1Apr20}.
From Eq.~\eqref{004019_25Jun20}, we have
\begin{align}
\Phi_1^\dagger \sigma^a \Phi_1 = v^2\left(f_1\f{x}{r}, f_1\f{y}{r}, f_1^2 c^2_{1/2} - h_1^2 s^2_{1/2}\right)
\end{align}
\begin{align}
\Phi_2^\dagger \sigma^a \Phi_2 = v^2\left(f_2\f{x}{r}, f_2\f{y}{r}, h_2^2 c^2_{1/2}  - f_2^2 s^2_{1/2}\right)
\end{align}
with $c_{1/2}\equiv \cos \f{\theta}{2}$ and $s_{1/2}\equiv \sin \f{\theta}{2}$,
and the unit vectors defined by Eq.~\eqref{080112_1Apr20} are
\begin{equation}
 n_1^a = \frac{f_1 h_1}{f_1^2 c_{1/2}^2+ h_1^2 s_{1/2}^2}
  \left( \f{x}{r}, \f{y}{r}, \f{f_1^2 c_{1/2}^2 - h_1^2 s_{1/2}^2}{f_1 h_1}\right),\label{214110_3Jul20}
\end{equation}
\begin{equation}
 n_2^a = \frac{f_2 h_2}{h_2^2 c_{1/2}^2 + f_2^2 s_{1/2}^2}
  \left( \f{x}{r}, \f{y}{r}, \f{h_2^2 c_{1/2}^2  - f_2^2 s_{1/2}^2}{f_2 h_2}\right).\label{214115_3Jul20}
\end{equation}
Note that $n_1^a$ and $n_2^a$ are not well-defined at $\theta=0$ and $\theta=\pi$, respectively
because the denominators vanish.

Furthermore, we require that the $U(1)_\mr{EM}$ is not broken everywhere, 
which leads to $n_1^a=n_2^a$ everywhere except for $\theta=0,\pi$.%
\footnote{Note that, although $n_1^a$ ($n^a_2$) is not defined on $\theta=0$ ($\theta=\pi$),
the $U(1)_\mr{EM}$ is not broken even on the $\theta=0,\pi$. 
It is broken only when both of $n^a_1$ and $n^a_2$ are well-defined and $n_1^a \neq n_2^a$.
}
We can explicitly confirm that this is true by constructing a full solution of the EOMs as in Refs.~\cite{Eto:2019hhf,Eto:2020hjb}.
This leads to two conditions for the profile functions $f_i, h_i$.
The expressions are rather complicated, so that we do not present them
although they are assumed implicitly in the following.

\begin{figure}[tbp]
 \begin{minipage}[c]{0.4\textwidth}
 \centering
 \includegraphics[width=0.5\textwidth]{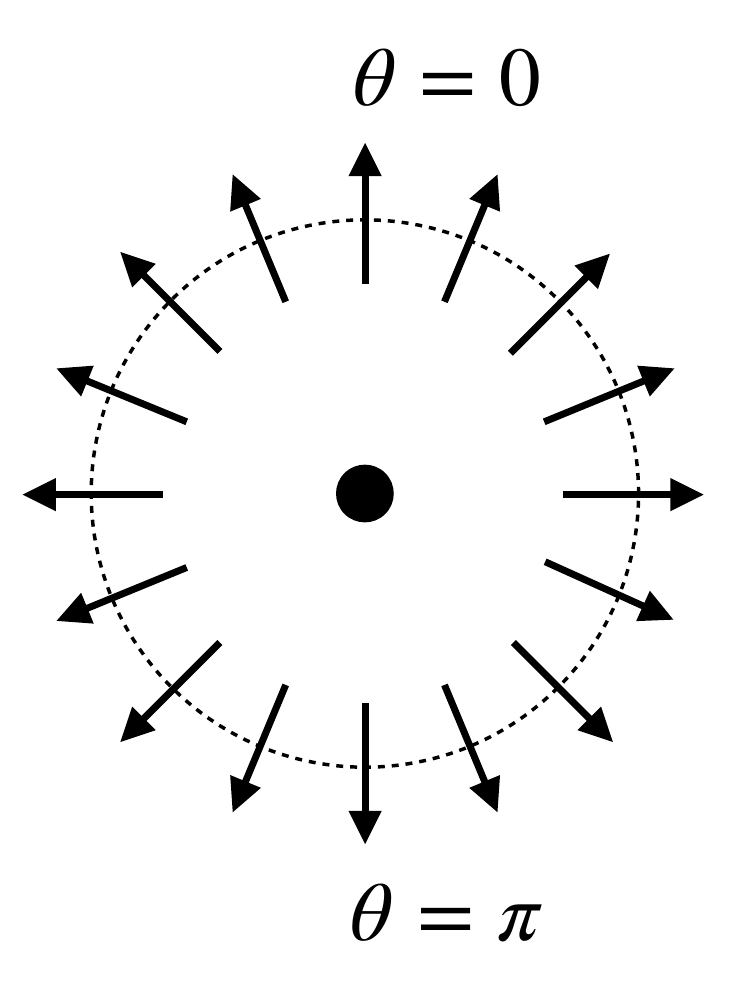}
  \subcaption{
  't Hooft-Polyakov monopole
  }
 \label{075619_30Jul20}
\end{minipage}
 \begin{minipage}[c]{0.4\textwidth}
 \centering
 \includegraphics[width=0.53\textwidth]{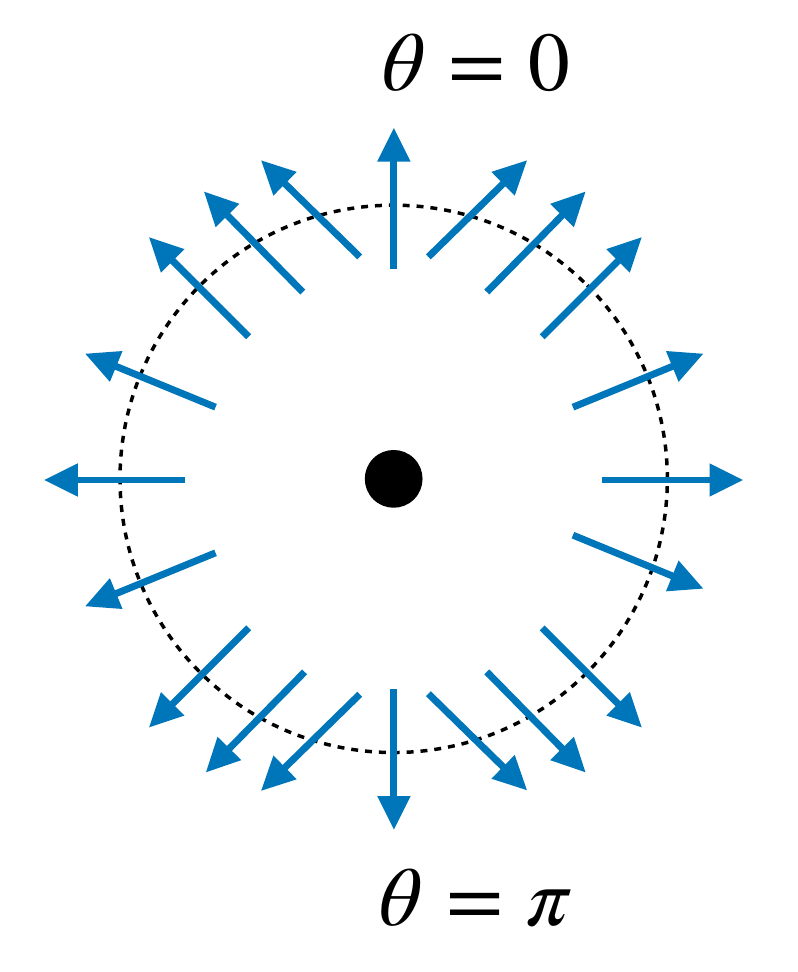}
  \subcaption{
  Nambu monopole in 2HDM
  }
\end{minipage}
 \label{223214_3Jul20}
 \caption{
 Structure of the unit vectors $n^a$. 
 (a): For the 't Hooft-Polyakov monopole, $n^a$ is defined by Eq.~\eqref{181300_26Jun20} and is the hedgehog structure.
 (b): For the Nambu monopole in 2HDM, $n^a$ is given by Eq.~\eqref{025136_19Jul20} and is not the hedgehog structure.
 It deforms around $\theta \sim 0$ and $\pi$.
}
 \label{103806_30Jul20}
 \end{figure}

We next consider the unit vector $n^a$ defined by Eq.~\eqref{140145_1Apr20}.
Substituting Eqs.~\eqref{214110_3Jul20} and \eqref{214115_3Jul20}, we obtain
\begin{equation}
 n^a = \frac{f_1 h_1 + f_2 h_2}{(f_1^2 + h_2^2)c_{1/2}^2 + (f_2^2 + h_1^2) s_{1/2}^2}
  \left(\f{x}{r}, \f{y}{r}, \f{(h_2^2 + f_1^2) c_{1/2}^2 - (h_1^2+f_2^2) s_{1/2}^2}{f_1 h_1 + f_2 h_2}\right),\label{223359_3Jul20}
\end{equation}
which is {\it regular and well-defined everywhere even at $\theta=0, \pi$}.
Indeed, $n^a = n^a_2=(0,0,1)$ for $\theta=0$ and $n^a =n_1^a=(0,0,-1)$ for $\theta=\pi$.
(Recall $f_1\to 0$ and $f_2\to 0$ for $\theta \to 0$ and $\pi$, respectively.)
As we move from the string core, $\rho \to \infty $, $n^a$ approaches to the hedgehog one as in the 't Hooft-Polyakov case,
\begin{equation}
 n^a = (\sin \theta \cos \varphi, \sin \theta \sin \varphi, \cos \theta)^t = \frac{x^a}{r}.\label{215607_3Jul20}
\end{equation}
On the other hand, it slightly deforms around $\theta\sim 0,\pi$ as is depicted by Fig.~\ref{103806_30Jul20}.

For later use, we rewrite $n^a$ at finite $\rho$, Eq.~\eqref{223359_3Jul20}, as
\begin{align}
 n^a =  \left(\sin\Theta \cos \varphi, \sin\Theta \sin\varphi, \cos \Theta\right)\label{025136_19Jul20}
\end{align}
where  $\Theta(\theta)$ is a ``deformed zenith angle,'' defined by 
\begin{align}
 \cos \Theta &\equiv \f{(h_2^2 + f_1^2)c_{1/2}^2  - (h_1^2 + f_2^2) s_{1/2}^2 }{(f_1^2 + h_2^2)c_{1/2}^2 + (f_2^2 + h_1^2) s_{1/2}^2} \\
\sin \Theta &\equiv \f{(f_1 h_1 + f_2 h_2 )}{(f_1^2 + h_2^2)c_{1/2}^2 + (f_2^2 + h_1^2) s_{1/2}^2} \f{\sqrt{x^2+y^2}}{r}.
\end{align}
Note $\cos ^2 \Theta + \sin ^2 \Theta=1$ since $|n^a|^2=1$,
and $ \Theta(0)= 0$, $\Theta(\pi) = \pi$.
In the matrix representation,
\begin{equation}
 n\equiv n^a \f{\sigma^a}{2} = \f{1}{2}
\begin{pmatrix}
\cos \Theta & \sin \Theta e^{ -i\varphi} \\ \sin\Theta e^{i\varphi} & -\cos \Theta
\end{pmatrix}.
\end{equation}

Because $n_1^a=n_2^a$ holds everywhere except for $\theta=0,\pi$,
we can define the field strengths of the electromagnetic and the $Z$ gauge bosons
by using Eqs.~\eqref{231040_24Jun20} and \eqref{004333_1May20} and the vector $n^a$, Eq.~\eqref{025136_19Jul20}.
The expressions are, however, rather complicated, and hence 
we here see just the asymptotic forms far from the string cores $\rho \to \infty$
by substituting the asymptotic forms Eq.~\eqref{200928_27May20} (or Eq.~\eqref{215607_3Jul20})
into Eqs.~\eqref{231040_24Jun20} and \eqref{004333_1May20}.
Then we obtain \cite{Eto:2019hhf}
\begin{align}
 F_{ij}^Z &=  \f{2\pi \cos \theta_\mr{W}}{g}  \f{z}{|z|}  \epsilon_{3ij}\delta (x) \delta (y),\label{232742_24Jun20}\\
 F_{ij}^\mr{EM} &=  \f{\sin \theta_\mr{W}}{g} \epsilon^{aij} \f{x^a}{r^3}\label{231737_24Jun20}.
\end{align}
From Eq.~\eqref{231737_24Jun20}, it is clear that there is a magnetic flux emanating from the origin in a spherical hedgehog form.
The total amount of the magnetic flux $\Phi_{\rm EM}$ can be calculated by integrating the flux density 
$B_i \equiv \f{1}{2}\epsilon_{ijk}F_{jk}^\mr{EM}$ as
\begin{equation}
 \Phi_{\rm EM} = \int d^3x~ \partial_i B_i =\f{4 \pi \sin \theta_\mr{W}}{g}.
\label{eq:EM_Sigma}
\end{equation}
In addition, from Eq.~\eqref{232742_24Jun20},
the $Z$-fluxes only exist on the $z$-axis as
\begin{eqnarray}
\Phi_Z\big|_{z > 0} &=& \int dx^2 \, F_{ij}^Z\big|_{z > 0} = \frac{2\pi\cos\theta_\mr{W}}{g}  = \Phi_Z^{(1,0)},\\
\Phi_Z\big|_{z < 0} &=& \int dx^2 \, F_{ij}^Z\big|_{z < 0} = - \frac{2\pi\cos\theta_\mr{W}}{g} = \Phi_Z^{(0,1)},
\end{eqnarray}
flowing on the positive and negative sides of the $z$-axes, respectively, from the origin.
These amounts of the $Z$-fluxes agree with ones of the $Z$ strings in Eq.~\eqref{eq:Z_fluxes}.
The total amount of the $Z$-fluxes flowing from the monopole at the origin can be calculated as
\begin{eqnarray}
\Phi_Z= \int dx^3\, \partial_i B_i^Z = \Phi_Z\big|_{z > 0} - \Phi_Z\big|_{z < 0} = \frac{4\pi \cos \theta_\mr{W}}{g},
\end{eqnarray}
with $B_i^Z \equiv \frac{1}{2}\epsilon_{ijk}F_{jk}^Z$.
Therefore, the total $Z$ flux agrees with the one of the Nambu monopole in the SM \eqref{204225_3Jul20}.
See Fig.~\ref{001942_1May20} for a schematic picture of the $Z$- and electromagnetic fluxes from the monopole.

It is worthwhile to demonstrate the topological current of $U(1)_a$ in the configuration.
The topological current, corresponding to the winding of the $U(1)_a$ phase of the Higgs field,
is defined by 
\begin{equation}
\mathcal{A}_i \equiv \epsilon_{ijk} \partial _j \mathcal{J}_k,\label{013151_11Mar20}
\end{equation}
\begin{equation}
\mathcal{J}_i \equiv -i ~\mr{tr} \left[H^\dagger D_i H  - (D_i H) ^\dagger H\right].
\end{equation}
Importantly, $\mathcal{A}_i$ is topologically conserved $\partial_i \mathcal{A}_i=0$, 
and independent of $z$ as seen by substituting Eq.~\eqref{200928_27May20}.
This indicates that not only the string parts but also the monopole itself at the origin has the topological charge of $U(1)_a$.

\subsection{Fiber bundle for Nambu monopole}
 \begin{figure}[tbp]
 \centering
 \includegraphics[width=0.5\textwidth]{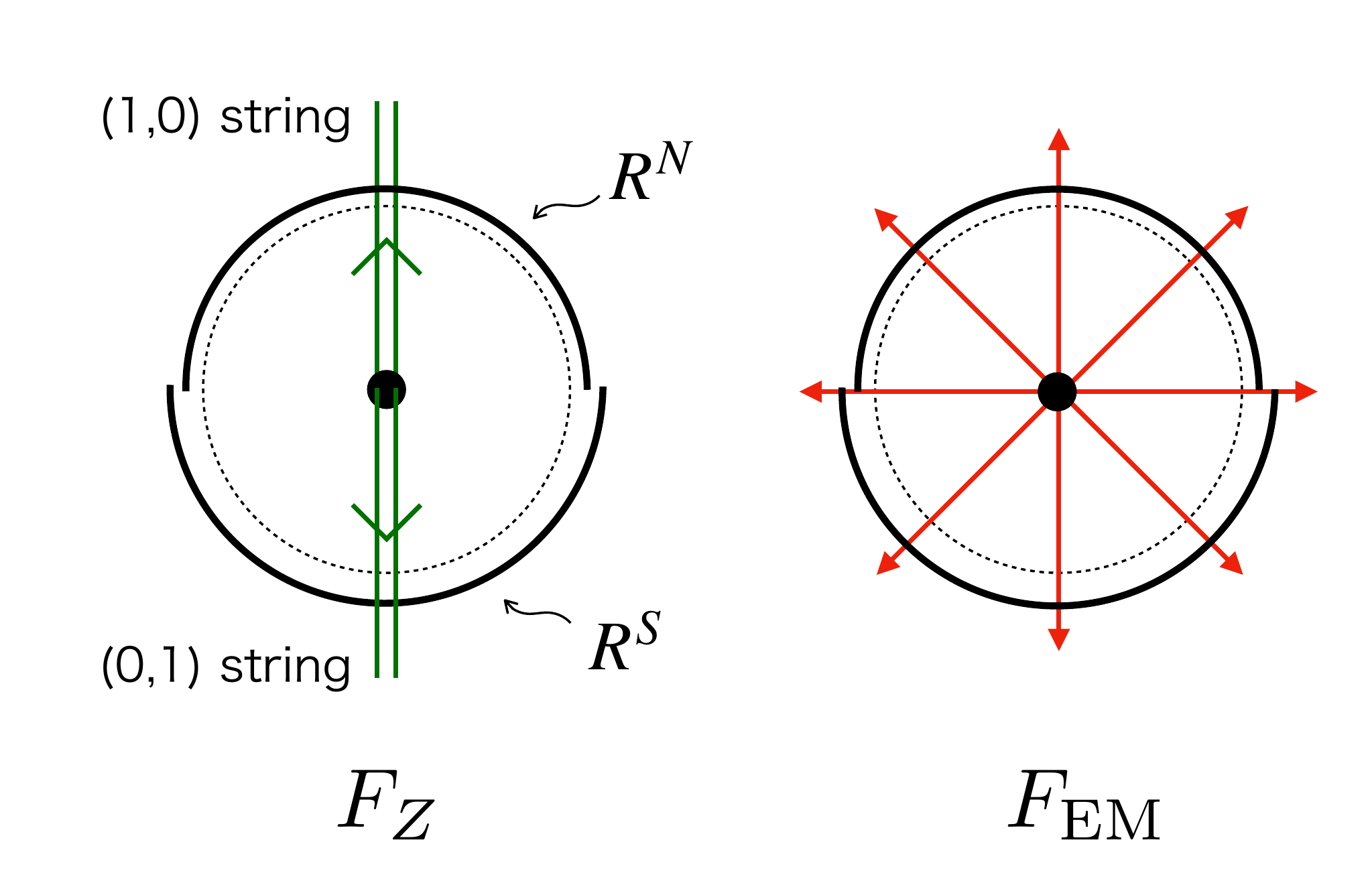}
  \caption{
  Schematic picture of the two patches $R^N$ and $R^S$ surrounding the Nambu monopole.
  The (1,0) and (0,1) string pass through $R^N$ and $R^S$, respectively.
  The electromagnetic flux, $\Phi_\mr{EM}=4 \pi \sin \theta_\mr{W}/g$ spreads radially from the monopole (right panel),
  while the $Z$ flux, $\Phi_Z=2\pi \cos \theta_\mr{W}/g$, is confined in each $Z$ string, $(1,0)$ or $(0,1)$ strings (left panel).
  }
  \label{001942_1May20}
 \end{figure}

We here discuss the topological structure of the electromagnetic field of the Nambu monopole.
Especially, we investigate a fiber bundle consisting of the base space $S^2$ surrounding the monopole and the fiber $U(1)_{\rm EM}$.

Importantly, since the strings pass through the base space $S^2$ at $\theta=0,\pi$ (see Fig.~\ref{001942_1May20}),
we cannot use the asymptotic form Eqs.~\eqref{200928_27May20}-\eqref{201042_27May20} on the sphere $S^2$,
which are singular on the north and the south poles.
Thus, we have to use the smeared one \eqref{004019_25Jun20}-\eqref{201042_27May20} without the line singularities.

We should stress that the unit vector $n^a$ \eqref{025136_19Jul20} is homotopically equivalent to the hedgehog one \eqref{215607_3Jul20},
{\it i.~e.~}, the map 
\begin{equation}
n^a(\theta , \varphi) : S^2=\{0\leq \theta \leq \pi, 0\leq \varphi < 2\pi\} \mapsto S^2 =\{n^a=(n^1,n^2,n^3) | (n^1)^2+(n^2)^2+(n^3)^2 =1\}
\end{equation}
has a winding number unity.
The target space $S^2$ corresponds to the order parameter space $SU(2)_W / U(1)_n$, 
where $U(1)_n$ is a subgroup of $SU(2)_W$ that leaves $n^a$ invariant.%
\footnote{
Although $U(1)_n$ keeps $n^a$ invariant, it does keep neither $\Phi_1$ nor $\Phi_2$ invariant,
thus $U(1)_n$ itself is not an unbroken subgroup.
}
Therefore, $n^a$ has the same topological structure as the 't Hooft-Polyakov monopole
(see Eq.~\eqref{104517_30Jul20} and texts around it),
and the Nambu monopole in 2HDM can be regarded as an embedding of the 't Hooft-Polyakov monopole into the $SU(2)_W$ sector accompanied by the $U(1)_Y$ flux.

To see it more closely, let us first consider the fiber bundle of the $U(1)_n$ subgroup.
Thanks to the fact that the unit vector $n^a$ in Eq.~\eqref{223359_3Jul20} or \eqref{025136_19Jul20} is regular everywhere,
we can define the $U(1)_n$ subgroup on  the whole sphere $S^2$ surrounding the monopole,
unlike the case of the Nambu monopole in the SM.
Thus, we should divide $S^2$ into two hemi-spheres $R^N$ and $R^S$ 
as in the case of the 't Hooft-Polyakov monopole (Sec.~\ref{154249_25Jun20}):
\begin{equation}
  R^N : 0 \leq \theta \leq \f{\pi}{2} , \h{2em}   R^S : \f{\pi}{2} \leq \theta \leq \pi ,
\end{equation}
which have an overlap region on the equator $\theta = \pi /2$.
What is different from the `t Hooft-Polyakov monopole is that we now have the $Z$ strings at $\theta=0$ and $\theta=\pi$,
and thus the strings pass through $R^N$ and $R^S$ (See Fig.\ref{001942_1May20}).
However, it is irrelevant  
because $n^a$ has no singularity 
even at the centers of the $Z$ strings.

Then, we introduce the $SU(2)_W$ gauge transformations $U^N$ and $U^S$ defined on each hemi-sphere.
\begin{align}
R^N : ~ U ^N(\theta , \varphi) &= e^{-i \sigma ^3 \varphi / 2} e^{-i \sigma ^2 \Theta / 2} e^{i \sigma ^3 \varphi / 2}
= \begin{pmatrix}
     \cos \f{\Theta}{2} & -\sin \f{\Theta}{2} e^{-i \varphi} \\
     \sin \f{\Theta}{2} e^{i \varphi}  & \cos \f{\Theta}{2}
 \end{pmatrix} \\
R^S : ~ U^S (\theta , \varphi) &= e^{-i \sigma ^3 \varphi / 2} e^{-i \sigma ^2 (\pi-\Theta) / 2} e^{i \sigma ^3 \varphi / 2} (-i \sigma^2)
= \begin{pmatrix}
    e^{-i \varphi} \cos \f{\Theta}{2} & -\sin \f{\Theta}{2}  \\
     \sin \f{\Theta}{2}  &  e^{i \varphi} \cos \f{\Theta}{2}
 \end{pmatrix}.
\end{align}
Note that we have used $\Theta$ instead of $\theta$.
The both transformations bring the unit vector $n^a$ into the uniform vector on each region:
\begin{align}
 U^N : ~ n^a\f{\sigma^a}{2} &\to \left(U^N\right)^\dagger n^a\f{\sigma^a}{2} U^N= \f{\sigma^3}{2}\\
 U^S : ~ n^a\f{\sigma^a}{2} &\to \left(U^S\right)^\dagger n^a\f{\sigma^a}{2} U^S= \f{\sigma^3}{2}.
\end{align}
Note that $U^N$ and $U^S$ are regular on each hemi-sphere $R^N$ and $R^S$.
In this gauge, the subgroup $U(1)_n$ is uniformly defined as the $\sigma^3$ subgroup of $SU(2)_W$ on the two hemispheres.

Similarly to Sec.~\ref{154249_25Jun20}, it follows from the single-valuedness of $n^a$ on $\theta = \pi/2$
that the transition function $(U^S)^{-1} U^N $ must leave $n^a$ invariant,
{\it i.~e.~}, $(U^S)^{-1} U^N \in U(1)_{n}$ (not $U(1)_\mr{EM}$).
Indeed, noting that $\Theta =\pi/2$ for $\theta = \pi/2$, we have
\begin{align}
 \left(U^S\right)^{-1} U^N= e^{ i \varphi \sigma^3},
\end{align}
which is the same as Eq.~\eqref{000509_1May20}
and has a winding number unity.


Thus, the first Chern number associated with $U(1)_n$ is obtained as
\begin{equation}
  \int_{S^2} c_1^{(n)}= \f{g}{2\pi} \int_{S^2} F^n = 1,\label{210834_30Jul20}
\end{equation}
with $F^n_{ij} \equiv n^a W_{ij}^a$ (not $F^Z_{ij}$ or $F^\mr{EM}_{ij}$).
Therefore, the topological structure of $U(1)_n (\subset SU(2)_W)$ is completely the same as
the unbroken subgroup of the 't Hooft-Polyakov monopole.

Let us turn into the topological structure of $U(1)_{\rm EM}$,
which is defined by the linear combination of $U(1)_n$ and $U(1)_Y$ (Eq.~\eqref{004333_1May20}).
Since the divergence of the $U(1)_Y$ flux is always zero,
the global structure of the fiber $U(1)_Y$ is trivial,
{\it i.e.}, the fiber bundle is a direct product: $S^2 \times U(1)_Y$.
Therefore, we can concentrate on the $U(1)_n$ part.
We consider the first Chern number associated with $U(1)_{\rm EM}$,
which is given by%
\footnote{
We choose a normalization factor for $c_1^\mr{(EM)}$ such that it takes an integer value under the integration.
}
\begin{equation}
  \int_{S^2}c_1^\mr{(EM)}
   \equiv \f{-g}{2\pi \sin\theta_W} \int_{S^2} F^\mr{EM} = \f{g}{2\pi} \int_{S^2} F^{n} = \int_{S^2}c_1^{(n)}= 1.\label{222627_5Jul20}
\end{equation}
This means that the magnetic charge
is quantized 
because $c_1^{(n)}$ is quantized as Eq.~\eqref{210834_30Jul20}.
Such a quantization originates from that 
the topological structure of $U(1)_n $ of this monopole has the same topology with 
that of the 't Hooft-Polyakov monopole (or the Wu-Yang monopole bundle).

Finally, let us consider the Dirac's quantization condition.
When a test particle with the $U(1)_Y$ hypercharge $Y$ and the weak isospin $T_3=1/2$ goes around on the equator $\theta = \pi/2$,
it receives an AB phase, given by
\begin{align}
 \theta_\mr{AB} &=  e Q \oint_{\theta = \pi/2} dx_i A_i  + g_Z  T_Z \oint_{\theta = \pi/2} dx_i Z_i\label{111623_30Jul20} 
\end{align}
where we have used $e= g \sin \theta_W $ and
\begin{equation}
 Q \equiv T_3 +\f{Y}{2}\label{154127_30Jul20}
\end{equation}
\begin{equation}
 T_Z \equiv T_3 - \sin^2 \theta_W Q.\label{154140_30Jul20}
\end{equation}
Note that, in Eq.~\eqref{111623_30Jul20}, 
there is also a contribution from the $Z$ flux confined inside the two $Z$ strings (the second term),
in addition to that from the electromagnetic gauge field (the first term).
Using the definitions of the gauge fields in Eqs.~\eqref{065506_1Apr20} and \eqref{005733_24Jun20}
and noting $n^a=(0,0,1)$ on each patch,
we obtain
\begin{align}
\theta_\mr{AB} &= \oint_{\theta= \pi/2} dx_i \left( -g T_3  W_i^3 + g' \f{Y}{2} Y_i  \right).\label{020323_1May20}
\end{align}

The second term in Eq.~\eqref{020323_1May20} vanishes because the $U(1)_Y$ fiber is trivial.
We can calculate the AB phases in the two patches, $R^N$ and $R^S$.
Noting that the difference of $W^3_i$ on the two patches is given by
\begin{equation}
\Delta W_i^3 = \f{-i}{g}\mr{Tr}\left( \sigma_3\left[ \left(U^S\right)^{-1} U^N\right]^\dagger \partial_i \left(U^S\right)^{-1} U^N \right),
\end{equation}
the condition for the single-valuedness of the wave function of the particle requires that 
the difference between the two patches,
\begin{align}
 \Delta \theta_{AB} &= g T_3 \oint_{\theta= \pi/2} dx_i~ \Delta W_i^3 
=4\pi T_3,\label{113848_30Jul20}
\end{align}
must be an integer multiple of $2\pi$:
\begin{align}
 &  4 \pi T_3 = 2\pi n ~ (n \in \mathbb{Z}) 
~ \Leftrightarrow~ T_3 = \f{n}{2},\label{021604_1May20}
\end{align}
which is the Dirac's quantization condition derived from the Nambu monoopole in 2HDM.
Recalling $T_3=1/2$,
this is automatically satisfied with $n=1$.
Remarkably, the Dirac quantization condition for the Nambu monopole in 2HDM does not ensure quantization of the electric charges 
because the fiber bundle of $U(1)_Y$ is trivial and the $U(1)_Y$ hypercharge cannot be quantized by the monopole. 

Before closing this section, we give some remarks.
In the SM, in the presence of the Nambu Monopole, 
the magnetic charge is not quantized
and the $U(1)_\mr{EM}$ and $Z$ gauge fields cannot have any topologically non-trivial structures.
This is because 
the electroweak symmetry is restored
and hence $U(1)_\mr{EM}$ cannot be defined at the center of the $Z$ string.
On the other hand, in 2HDM, 
the magnetic charge is quantized as in Eq,~\eqref{222627_5Jul20} 
because the topological charge (first Chern number) $c_1^{(n)}$ is quantized.
This originates from that the Nambu monopole in 2HDM is an embedding of the 't Hooft-Polyakov monopole into the $SU(2)_W$ sector.
What makes the embedding possible is that 
$n^a$ defined by Eq.~\eqref{140145_1Apr20} is regular and well-defined everywhere on the infinitely large sphere $S^2$ surrounding the monopole.
The regularity means that the electroweak symmetry is {\it NOT} restored on the whole sphere, even on the cores of the $Z$ strings.
In this sense, the Nambu monopole in 2HDM is indeed a true magnetic monopole although it is attached with the strings.
This topological argument is still valid even when the $U(1)_a$ and $(\mathbb{Z}_2)_\mr{C}$ symmetries are explicitly broken in the potential
because the breaking effects do not affect the topology of the 't Hooft-Polyakov embedding in $SU(2)_W \to U(1)_n$.
However, the stability is not ensured anymore
in the presence of such breaking terms, 
because we have no non-trivial second homotopy group $\pi_2$ in the full symmetry breaking $SU(2)_W \times U(1)_Y \to U(1)_\mr{EM}$.
The reader should not confuse the topological property of this monopole with the topological stability.
To stabilize the monopole, we need non-trivial $\pi_0$ and $\pi_1$ associated with the $(\mathbb{Z}_2)_\mr{C}$ and $U(1)_a$ symmetries as studied in Ref.~\cite{Eto:2019hhf}.
See Ref.~\cite{Eto:2020hjb} for the unstable case without $(\mathbb{Z}_2)_\mr{C}$.

\section{Dyon in 2HDM}
\label{044406_30Jul20}
Similarly to the electroweak dyon in the SM studied in Ref.~\cite{Vachaspati:1994xe}, we can consider a dyon configuration in 2HDM.
To do this, we consider the gauge moduli of the Nambu monopole described by Eqs.~\eqref{200928_27May20},\eqref{201019_27May20} and \eqref{201042_27May20}.
Hereafter, we denote the Nambu monopole configuration in 2HDM by $\overline{H}$, $\overline{W}^a_\mu$ and $\overline{Y}_\mu$.
The $SU(2)_W$ adjoint vector $n^a$ defined in Eq.~\eqref{140145_1Apr20}
can be rewritten in terms of the $2\times 2$ matrix $H$ as
\begin{equation}
n^a= \f{\mr{tr}\left(\sigma^3 H^\dagger \sigma^a H \right)}{\mr{tr}\left(H^\dagger H \right)},
\end{equation}
which satisfies
\begin{equation}
 \left(\overline{D}_\mu n\right)^a =0\label{234556_3Jun20}
\end{equation}
at large distances from the strings, $\rho \to \infty$, with $\overline{D}_\mu$ the covariant derivative consisting of $\overline{W}_\mu, \overline{Y}_\mu$.
In addition, the following identity holds:
\begin{equation}
 \overline{H}\sigma^3 + n^a \sigma^a \overline{H}=0,
\end{equation}
which means that $\overline{H}$ is invariant under the $U(1)_\mr{EM}$ gauge transformation:\footnote{
The $U(1)_\mr{EM}$ transformation is defined as a combination of the $U(1)_n$ and $U(1)_Y$ transformations.}
\begin{equation}
\exp \left(- i \alpha n^a \f{\sigma_a}{2}\right)\overline{H} \exp\left(-i \alpha \f{\sigma_3}{2}\right) =\overline{H}
\end{equation}
with $\alpha$ being an arbitrary function.

Due to the $U(1)_\mr{EM}$ symmetry, the monopole has a modulus (zero mode) under the following transformation:
\begin{align}
\overline{H} & \to \exp \left[- i\alpha n^a \f{\sigma_a}{2}\right]\overline{H} \exp\left[-i \alpha \f{\sigma_3}{2}\right]~ (= \overline{H})\label{031327_4Jun20} \\
g \overline{W} _\mu & \to  \exp\left[- i\alpha n^a \f{\sigma_a}{2}\right]\left( g \overline{W}_\mu - i \partial_\mu\right)  \exp\left[ i\alpha n^a \f{\sigma_a}{2}\right] \\
g' \overline{Y} _\mu & \to   g' \overline{Y}_\mu - \exp\left[i  \alpha\right] i \partial_\mu \exp\left[- i\alpha\right]~ (= g' \overline{Y}_\mu), \label{172230_6Jul20}
\end{align}
for constant $\alpha \in [0,2\pi)$.

Let us make a dyon configuration from the Nambu monopole.
In the following, we are interested only in the asypmtotic forms at large distances from the string cores $\rho \to \infty$.
If one wants to obtain a true configuration, it is necessary to solve the full EOMs.
The asypmtotic form of the dyon configuration is obtained by giving the monopole a time-dependent excitation using a function $\gamma(x)=\gamma(t,\bm{x})$,
\begin{equation}
 H= \overline{H},\label{025324_4Jun20}
\end{equation}
\begin{equation}
 g W_\mu ^a = g \overline{W}^a_\mu - \delta_{\mu 0}\, \left(\overline{D}_\mu(n \gamma (x))\right)^a= g \overline{W}^a_\mu - \delta_{\mu 0}\, n^a \dot{\gamma}\label{221235_30Jul20}
\end{equation}
\begin{equation}
 g' Y_\mu = g' \overline{Y}_\mu + \delta_{\mu 0}\, \partial_\mu \gamma(x)= g' \overline{Y}_\mu + \delta_{\mu 0}\, \dot{\gamma},\label{025338_4Jun20}
\end{equation}
where $\dot{\gamma}= \partial_t \gamma$ and we have used Eq.~\eqref{234556_3Jun20} in Eq.~\eqref{221235_30Jul20}.
This excitation is regarded as acting the $U(1)_\mr{EM}$ transformation Eqs.~\eqref{031327_4Jun20}-\eqref{172230_6Jul20} only for the time components with $\alpha \to \gamma(x)$.
Due to this, the field strengths are changed from the ones in the Nambu monopole.
The difference is given as
\begin{equation}
\delta W_{i0}^a = - \f{\left(\overline{D}_i (n \dot{\gamma})\right)^a}{g} = - \f{n ^a \partial_i  \dot{\gamma}}{g}\label{024116_28May20}
\end{equation}
\begin{equation}
\delta Y_{i0} = + \f{\partial_i \dot{\gamma}}{g'}\label{024126_28May20}
\end{equation}
and $\delta W_{ij}=\delta Y_{ij}=0$.
For the electromagnetic and the $Z$ field strengths,
\begin{equation}
\delta F_{i0}^\mr{EM} =   \f{\sqrt{g^2 + g'^2}}{ gg'}   \partial_i \dot{\gamma},\label{031158_4Jun20}
\end{equation}
and the others do not change.

Let us obtain $\gamma$ by solving the EOMs at large distances from the string core, $\rho \to \infty$.
For the ansatz in Eqs.~\eqref{025324_4Jun20}-\eqref{025338_4Jun20}, 
the EOMs reduce to the ordinary Maxwell's equations for the electric field:
\begin{equation}
\partial_0 F_{i0}^\mr{EM} =0,\h{2em} \partial_i F_{0i}^\mr{EM} =0,
\end{equation}
which lead to 
\begin{equation}
 \partial_i \partial_i \dot{\gamma}= \partial_0 \partial_i \dot{\gamma} =0.\label{025851_4Jun20}
\end{equation}
Note that the other components of the fields do not receive back reactions from $\gamma$ 
because the massive gauge components exponentially decay at large distances and the Higgs field does not couple to $\gamma$.
Therefore, the other EOMs are satisfied independently of $\gamma$.%
\footnote{
Ofcourse they are non-linearly coupled around the cores of the dyon and the string.
}

From Eq.~\eqref{025851_4Jun20}, we have
\begin{equation}
 \dot{\gamma} = c_0(t) - \f{\Lambda}{4 \pi r}
\end{equation}
with $\Lambda$ an arbitrary constant and $c_0(t)$ an arbitrary function of $t$.
By gauge fixing condition (see \eqref{022901_28May20} in App.~\ref{030932_4Jun20}), we impose $\ddot{\gamma}=0$ and hence $\dot{ c_0}(t)=0$.
We thus obtain a general solution of $\gamma$ as
\begin{equation}
 \gamma =  \left(c_0-\f{\Lambda}{4 \pi r} \right)t + \text{const.}\label{031401_4Jun20}
\end{equation}

Substituting this into Eqs.~\eqref{031158_4Jun20}, we then get
\begin{equation}
 E _i = F_{0i}^\mr{EM} = \Lambda  \f{\sqrt{g^2 + g'^2}}{ 4 \pi gg'} \f{x_i}{r^3} = \f{\Lambda}{e} \f{1}{4 \pi} \f{x_i}{r^3}.
\end{equation}
Therefore, the configuration in Eqs.~\eqref{025324_4Jun20}-\eqref{025338_4Jun20} with Eq.~\eqref{031401_4Jun20} indeed describes a dyon configuration with the electric charge $q=\Lambda /e$.

So far, the electric charge $q$ is arbitrary and continuous.
We finally discuss the charge quantization condition for the dyon.
Let us consider a pair of dyons
with the magnetic and electric charges $(q_M,q_1)$ and $(q_M,q_2)$ with $q_M =4\pi \sin^2\theta_W/e$.
From the quantization condition for the angular momentum in quantum mechanics, 
we can derive the condition (see Ref.\cite{Vachaspati:1994xe}) as \footnote{
Note that the analysis of the angular momentum around a pair of dyons in Ref.\cite{Vachaspati:1994xe}
cannot be justified in the SM
because there is a singular point in the $Z$ string where the electromagnetic field 
cannot be defined as stated in Sec.~\ref{061521_4Jun20}.
However, we can apply the result to the case of the dyons in 2HDM,
in which the electromagnetic field is well-defined everywhere except for the center of the monopole or dyon.
}
\begin{equation}
 q_1 - q_2 = n e, \h{2em} n \in \mathbb{Z}.
\end{equation}
If we use this condition for a pair of the dyon $(q_1=q)$ and the monopole $(q_2=0)$,
we obtain
\begin{equation}
 q = n e, \h{2em} n \in \mathbb{Z},
\end{equation}
and thus $\Lambda$ in Eq.~\eqref{031401_4Jun20} must be quantized as $\Lambda = n e^2$.
Therefore, the electric charge of the dyon must be a multiple by the minimal charge $e$.

\section{Conclusion and Discussion}
\label{212143_21Jul20}
We have studied topological properties of the Nambu monopole in 2HDM.
In the SM, the Nambu monopole has no topological structures
because the $U(1)_\mr{EM}$ gauge field cannot be defined
on the center of the $Z$ string, where 
the electroweak symmetry is restored.
Thus, a  $U(1)_\mr{EM}$  fiber bundle
over a spatial surface $S^2$ 
surrounding the monopole 
is topologically trivial: $U(1)_\mr{EM}\times S^2$. 
This is different from 
the 't Hooft-Polyakov monopole 
which has the same structure with 
the Wu-Yang description of the Dirac monopole for which 
$U(1)_\mr{EM}$ is Hopf fibered over 
$S^2$ with the total space $S^3$.
As a result of the trivial topology, the magnetic charge is not topologically quantized for Nambu monopole in the SM.
On the other hand, the monopole in 2HDM has the same topological structure as the 't Hooft-Polyakov monopole
thanks to a fact that $ U(1)_n$ and hence $ U(1)_\mr{EM}$ are well-defined everywhere (except for the center of the monopole),
implying that the electroweak gauge symmetry is broken everywhere 
including the string cores.
Concentrating on the $SU(2)_W\to U(1)_n$ sector, the Nambu monopole can be regarded as an embedding of the 't Hooft-Polyakov monopole,
and the fiber bundle of $U(1)_n$ on the base space $S^2$ surrounding the monopole is non-trivial: $S^3$,
as in the 't Hooft-Polyakov and Wu-Yang cases.
Consequently, the $U(1)_\mr{EM}$ gauge field has also a non-trivial topology on $S^2$ 
and the electromagnetic flux is fractionally quantized.
We have shown that
the Dirac's quantization condition, \eqref{021604_1May20}, 
always holds for test particles charged under the electroweak gauge symmetry

We have also shown the dyon configuration in 2HDM.
A stationary time-dependent excitation makes the Nambu monopole a dyon, having both the magnetic and electric charges.
Thus the dyon behaves as a source for the electric and magnetic fields and the $Z$ magnetic flux.
Considering quantum effects, the electric charge of the dyon is quantized by the minimal electric charge $e=g\sin \theta_W$.

Before closing this paper, some discussions are addressed here.
It is known that 
the CP-violating $\theta$ term: $g^2 \theta F \tilde{F} /32 \pi^2$ in the Lagrangian changes the 't Hooft-Polyakov monopole 
into a dyon with a fractional electric charge $q= (n - \theta/2\pi)e$,
which is called the Witten effect \cite{Witten:1979ey}.
Let us discuss the Witten effect in 2HDM.
When the couplings between the gauge fields and the SM fermions are neglected, 
the $\theta$ term of $SU(2)_W$ gauge sector would provide the same effect on the Nambu monopole in 2HDM.%
\footnote{The Witten effect for the Nambu monopole in the SM was argued in Ref.~\cite{Vachaspati:1994xe}.}
The couplings to the SM fermions make the $\theta$ term unphysical since it can be eliminated by redefinition of the fermions and the Higgs doublets.
However, if the theory has any explicit CP-violating parameters, they provide similar effects
through higher-order loop corrections after integrating out perturbations around the monopole configuration \cite{Witten:1979ey,Callan:1982au},
resulting in the Nambu monopole with a fractional electric charge.
In fact 2HDM with the fermions in general admits several CP violating parameters including the Cabibo-Kobayashi-Maskawa phase;
For instance, parameters in the Higgs potential, CP-violating Higgs VEVs and 
additional Yukawa couplings with the fermions.
How the induced fractional charge depends on the parameters is non-trivial but is calculable in principle.
The effects could be useful to probe the CP violation in 2HDM using the Nambu monopole or dyon.

In this paper, we have considered a single Nambu monopole or dyon configuration with the two strings.
In practice, there can be also an anti-monopole or anti-dyon on the string.
Since the electric charges of the dyon and the anti-dyon are arbitrary,
we can assign electric charges of the same sign for the pair,
leading to electrically repulsive force between them.
This can change the fate of the dyons and monopoles in the early universe as follows.
In realistic 2HDM, the $U(1)_a$ symmetry should be explicitly broken to give a mass for the CP-odd Higgs boson,
which results in domain walls (or membranes) attached to the strings \cite{Eto:2018hhg}.
The strings pulled by the wall tensions collide to each other and reconnect,
forming small string loops stretched by walls inside the loop.
Such small loops immediately shrink and annihilate.
However, if a string loop has a dyon pair with electric charges of the same sign,
the dyons receive electrically repulsive force,
preventing the loop to shrink at a ceratain loop size.
Thus, the string loop with the dyon pair behaves as a long-lived particle
and can be abundant in the present universe.
A quantitative discussion requires further studies, which we leave as the future work.

\section*{Acknowledgements}
We would like to thank Masafumi Kurachi for discussion in the early stage of this work.
We also thank Yoshihiko Abe, Ryuichiro Kitano, Masatoshi Yamada and Wen Yin for useful discussions. 
%
The work is supported in part by 
JSPS Grant-in-Aid for Scientific Research 
(KAKENHI Grant No. 
No. JP19K03839 (M. E.), 
No. JP18J22733 (Y. H.), 
No. JP18H01217 (M. N.)), 
and also by MEXT KAKENHI Grant-in-Aid for Scientific Research on Innovative Areas, 
%
%
Discrete Geometric Analysis for Materials Design, 
No. JP17H06462 (M. E.) 
from the MEXT of Japan.

\appendix
\section{Fiertz identities}
\label{044424_30Jul20}
In this Appendix, we summarize the Fiertz identities.
For $SU(N)$ generators $T^a$ such that
\begin{equation}
 \mathrm{Tr} (T^a T^b) = \f{1}{2}\delta^{ab},
\end{equation}
the following relations hold:
\begin{equation}
 \mathrm{Tr}(T^a X) \mathrm{Tr}(T^a Y) = \f{1}{2} \mathrm{Tr}(XY) - \f{1}{2N} \mathrm {Tr}(X) \mathrm{Tr}(Y)
\end{equation}
\begin{equation}
 \mathrm{Tr}(T^a X T^a Y) = \f{1}{2} \mathrm{Tr}(X) \mathrm{Tr}(Y) - \f{1}{2N}  \mathrm{Tr}(XY)\label{142651_1Apr20}
\end{equation}
with arbitrary $N\times N$ matricies $X,Y$.

Using these realtions, we have
\begin{align}
 (\Phi^\dagger \sigma^a \Phi) (\Phi^\dagger \sigma^a \Phi) &= \mr{Tr} \left[\sigma^a \Phi \Phi^\dagger \sigma^a \Phi \Phi^\dagger\right] \nonumber \\
&= 2 \mr{Tr} \left(\Phi \Phi^\dagger\right)\mr{Tr} \left(\Phi \Phi^\dagger\right) - \mr{Tr}\left(\Phi \Phi^\dagger \Phi \Phi^\dagger\right) \nonumber \\
&= (\Phi ^\dagger \Phi)^2 .
\end{align}
Other usuful relations also can be derived (see Ref.~\cite{Nambu:1977ag}).

\section{Background gauge condition}
\label{030932_4Jun20}
We here consider perturbations around a background field configuration in the $SU(2)_W \times U(1)_Y$ gauge theory.
Due to the gauge symmetry, perturbations in the direction of the gauge orbit should be identified
and any physical perturbations should be orthnormal to the gauge orbit.
This is called the background gauge condition.
Let $\delta W_\mu$, $\delta Y_\mu$ be the perturbations around the background fields $\overline{W}_\mu$, $\overline{Y}_\mu$.
The orthnormality reqires that the following conditions holds:
\begin{equation}
 \langle \delta W_\mu|\delta_\eta W_\mu \rangle =0 
\end{equation}
\begin{equation}
 \langle \delta Y_\mu|\delta_\chi Y_\mu \rangle =0 ,
\end{equation}
where $\delta_\eta W_\mu$ and $\delta_\chi Y_\mu$ are the gauge transformations
with the gauge functions $\eta^a (x)$ and $\chi(x)$, respectively.

Using $\delta_\eta W_\mu = \overline{D}_\mu \eta$ and $\delta_\eta Y_\mu = \partial_\mu \chi$,
these conditions can be rewritten as
\begin{equation}
 \int d^4 x ~\mr{Tr}\left(\delta W_\mu \overline{D}^\mu \eta \right) =0
\end{equation}
\begin{equation}
 \int d^4 x ~\delta Y_\mu \partial^\mu \chi =0
\end{equation}
for any $\eta(x)$ and $\chi(x)$.
By integrating them by parts, we obtain
\begin{equation}
\overline{D}^\mu\delta W_\mu = \partial^\mu \delta Y_\mu=0,\label{022901_28May20}
\end{equation}
which are called the background gauge conditions.

\section{Singular gauge for Nambu monopole in 2HDM}
\label{044644_30Jul20}
Here, we derive the Dirac's quantization condition for the Nambu monopole in 2HDM in a singular gauge.
We start from Eqs.~\eqref{004019_25Jun20}-\eqref{195520_18Jul20} again.
Recall the vector $n^a$ is Eq.~\eqref{025136_19Jul20}.

By performing the following gauge transformation
\begin{align}
 U &= e^{i \sigma ^3 \varphi / 2} e^{i \sigma ^2 \Theta / 2} e^{-i \sigma ^3 \varphi / 2} 
 = \begin{pmatrix}
     \cos \f{\Theta}{2} & -\sin \f{\Theta}{2} e^{-i \varphi} \\
     \sin \f{\Theta}{2} e^{i \varphi}  & \cos \f{\Theta}{2}
    \end{pmatrix},
\end{align}
which is singular only at $\theta= \pi$,
$n^a$ transforms as
\begin{equation}
 n \to  U^\dagger n U =\f{\sigma^3}{2},
\end{equation}
or equivalently, $n^a \to \left(0,0,1\right)$.
Under this transformation, the two doublets transform as
\begin{equation}
  \Phi_1 \to U^\dagger \Phi_1 = 
\begin{pmatrix}
e^{-i\varphi}\left(f_1 c_{1/2} \cos \f{\Theta}{2} + h_1 s_{1/2} \sin \f{\Theta}{2}\right) \\ 
h_1 s_{1/2} \cos \f{\Theta}{2} - f_1 c_{1/2} \sin \f{\Theta}{2}
\end{pmatrix}\label{002811_10Apr20}
\end{equation}
\begin{equation}
  \Phi_2 \to U^\dagger \Phi_2 = 
\begin{pmatrix}
 \left( h_2 c_{1/2} \cos \f{\Theta}{2} + f_2 s_{1/2} \sin \f{\Theta}{2}\right) \\ 
 e^{i \varphi} \left( f_2 s_{1/2} \cos \f{\Theta}{2} - h_2 c_{1/2} \sin \f{\Theta}{2} \right)
\end{pmatrix},\label{002823_10Apr20}
\end{equation}
with $s_{1/2} = \sin\f{\theta}{2} $ and $c_{1/2}= \cos \f{\theta}{2}$.
Eq.~\eqref{002811_10Apr20} is singular only at $\theta=\pi$ 
while Eq.~\eqref{002823_10Apr20} is regular everywhere.

This singular transformation gives also a singularity on $\theta= \pi$ to the $SU(2)_W$ gauge field as
\begin{align}
  g W_i &\to U^\dagger \left( g W_i - i \partial_i\right)U \nonumber \\
 &=- \cos^2 \theta_W \f{\sigma_3}{2} j(\theta)\cos \theta \partial_i \varphi 
+ U^\dagger \left( \f{\sigma_a}{2}\epsilon^{iab}\f{x^b}{r^2} - i\partial_i\right)U \nonumber \\
 &= -\cos^2 \theta_W \f{\sigma_3}{2}  j(\theta)\cos \theta \partial_i \varphi  
+ \f{\sigma_3}{2} \f{1- \cos \theta}{r \sin \theta} \hat{\varphi}_i + \mathcal{O}(\theta - \pi),\label{174541_23Jul20}
\end{align}
where the last term denotes some regular terms that vanish at $\theta \to \pi$.
This expression \eqref{174541_23Jul20} has a singularity at $\theta = \pi$.
In this gauge, the electromagnetic gauge field is (note that $Y_i \to 0$ at $\theta \to \pi$ due to $k(\pi)=0$)
\begin{align}
  A_i &= -\sin\theta_W W_i^3 + \cos\theta_W Y_i \nonumber \\
 & =  \f{\sin\theta_W}{g} (1- \cos \theta) \partial_i\varphi + \mathcal{O}(\theta - \pi) ,
\end{align}
and we have
\begin{equation}
 \partial_{[i} A_{j]}= \f{2\pi \sin\theta_W}{g} \left(1-\f{z}{|z|}\right) \delta(x)\delta(y).
\end{equation}
On the other hand, the $Z$ gauge field is
\begin{align}
  Z_i &= -\cos\theta_W W_i^3 - \sin\theta_W Y_i \nonumber \\
 & =  \f{\cos\theta_W}{gr} \f{1- \cos \theta}{\sin \theta} \hat{\varphi}_i + \mathcal{O}(\theta - \pi) \nonumber \\
 & =  \f{\cos\theta_W}{g} (1- \cos \theta) \partial_i\varphi + \mathcal{O}(\theta - \pi) ,
\end{align}
which also has a line singularity at $\theta = \pi$.

Let us consider a test particle moving around the line singularity, {\it i.~e.},  with $\varphi $ from 0 to $2\pi$ and $\theta \sim \pi$,
 and calculate the AB phase $\theta_{AB}$:
\begin{align}
 \exp[i\theta_{AB}]&= \exp \left[i\oint_{\theta\approx \pi} dx_i \left(e Q A_i + g_Z T_Z Z_i  \right) \right],\label{015224_24Apr20}
\end{align}
where $Q$ and $ T_Z$ are defined in Eqs.~\eqref{154127_30Jul20} and \eqref{154140_30Jul20}.

Eq.~\eqref{015224_24Apr20} reads
\begin{align}
 \eqref{015224_24Apr20}
 %
 %
 &= \exp \left[i\oint_{\theta \approx \pi} dx_i \left( -g T_3  W_i^3 + g' \f{Y}{2} Y_i  \right) \right] \nonumber \\
 &= \exp \left[ -i T_3 \oint_{\theta \approx \pi} dx_i (1- \cos \theta) \partial_i\varphi \right] \nonumber \\
 &= \exp \left(- 4 \pi i T_3 \right)
\end{align}

Since the line singularity is gauge artifact,
it must not be observed, and we thus obtain the following condition
\begin{align}
 &  4 \pi T_3 = 2\pi n ~ (n \in \mathbb{Z})
 ~ \Leftrightarrow ~ T_3 = \f{n}{2},
\end{align}
which is the same condition as Eq.~\eqref{021604_1May20} derived from the argument on the fiber bundle.
Note that we have to take into account the line singularities not only in $A_i$ but also in $Z_i$.

\bibliographystyle{apsrev4-1}
\bibliography{./references}

\end{document}